\documentclass[%
aip,
amsmath,amssymb,
amsfonts,amsthm,amsopn, 
reprint,
]{revtex4-1} 

\usepackage{dcolumn}
\usepackage{bm}

\usepackage[utf8]{inputenc}
\usepackage[T1]{fontenc}
\usepackage{mathptmx}

\usepackage{siunitx}
\usepackage{hyperref}

\usepackage{xcolor}
\usepackage{color,soul}

\usepackage[absolute,overlay]{textpos}
\usepackage{graphicx,tabularx}
\usepackage{array}
\usepackage{subcaption}

\newcommand{\beginsupplement}{%
	\setcounter{table}{0}
	\renewcommand{\thetable}{S\arabic{table}}%
	\setcounter{figure}{0}
	\renewcommand{\thefigure}{S\arabic{figure}}%
	\setcounter{equation}{0}
	\renewcommand{\theequation}{s\arabic{equation}}%
}

\begin{document}
	\preprint{AIP/123-QED}
	
	\title{Time-scale coupling in hydrogen- and van der Waals-bonded liquids}
	\author{Lisa A. Roed} 
	\author{Jeppe C. Dyre} 
	\author{Kristine Niss} 
	\author{Tina Hecksher} 
	\author{Birte Riechers} 
	\email{birteri@ruc.dk}
	\affiliation{%
		Glass and Time, IMFUFA, Department of Science and Environment, Roskilde University, P.O. Box 260, DK-4000 Roskilde, Denmark
	}%
	\date{\today}
	
	\begin{abstract}
		The coupling behavior of time scales of structural relaxation is investigated on the basis of five different response functions for 1,2,6-hexanetriol, a hydrogen-bonded liquid with a minor secondary contribution, and 2,6,10,15,19,23-hexamethyl-tetracosane (squalane), a van der Waals bonded liquid with a prominent secondary relaxation process. Time scales of structural relaxation are derived as inverse peak frequencies for each investigated response function. For 1,2,6-hexanetriol, the time-scale indices are  temperature-independent, while a decoupling of time scales is observed for squalane in accordance with literature. An alternative evaluation approach is made on the squalane data, extracting time scales from the terminal relaxation mode instead of the peak position, and in this case temperature-independent coupling is also found for squalane, despite its strong secondary relaxation contribution. Interestingly, the very same ordering of response-function-specific time scales is observed for these two liquids, which is also consistent with the observation made for simple van der Waals bonded liquids reported previously [Jakobsen \textit{et al.}, J. Chem. Phys. \textbf{136}, 081102 (2012)]. This time-scale ordering is based on the following response functions, from fast to slow dynamics: shear modulus, bulk modulus, dielectric permittivity, longitudinal thermal expansivity coefficient, and longitudinal specific heat. These findings indicate a general relation between the time scales of different response functions and, as inter-molecuar interactions apparently play a subordinate role, suggest a rather generic nature of the process of structural relaxation.
		
	\end{abstract}

	
	
	\maketitle
	\section[Introduction]{Introduction}
	
	The dynamics of structural relaxation in amorphous matter show tremendous temperature-imposed changes, resulting in structural response time scales, $\tau$, spanning up to 15 orders of magnitude \cite{angell1985strong,dyr06,FitParabol2009}. In the liquid regime, structural relaxation times are in the picosecond range, whereas deeply supercooled liquids for which the equilibrium state can barely be reached have response time scales of $\tau \approx$~\SI{1000}{\second}.
	
	From an experimental point of view, the time scale of structural response can be assessed in multiple ways, e.g. by dielectric and mechanical spectroscopy but also based on light-scattering \cite{pabst2017}, calorimetric \cite{SHOIFET2015}, or aging measurements \cite{AgingGLY2019}. Time scales can be determined from the loss-peak position in spectral data, from auto-correlation functions based on time-domain data, or from the terminal mode reflecting the slowest spectral response such as the Maxwell relaxation time, $\tau_{M}$ \cite{harrison1976dynamic}. Interestingly, time scales obtained from different response functions are not identical, even if determined the same way, e.g., via the spectral loss-peak position. The relation of time scales for two different response functions $X_1$ and $X_2$ is conveniently quantified by the following \textit{time-scale index}, $\log \tau_{X_1}(T) - \log \tau_{X_2}(T)$.
	
	If both sets of time scales have the same dependence on temperature, $T$, the time-scale index becomes temperature-independent. This \textit{coupling} of time scales is observed for various pairs of response functions for a range of molecular glass formers \cite{vdw2012GT}. Several experimental studies report this behavior on the basis of data for the dielectric permittivity, $\epsilon$, and the dc conductivity, $\sigma$, for phenyl salicylate, propylene carbonate, phenolphthaleine-dimethyl-ether, and various mono-alcohols \cite{Stickel1996}, the viscosity, $\eta$, for o-terphenyl, glycerol, and propylene carbonate \cite{Hansen1997,Kremer2018}, calorimetric data on 5-polyphenyl-4-ether (5PPE) \cite{SHOIFET2015}, and the dynamic shear modulus, $G(\omega)$, for 5PPE, 1,3-butandiol, and other molecular liquids  \cite{GT1994comp, Manda2001, 7liqs2005GT}. Constant time-scale indices have also been reported for shear- and bulk-mechanical data on 5PPE, tetramethyl-tetraphenyl-trisiloxolane (DC704), 1,2,6-hexanetriol, and more \cite{GT1994HEX, GT2013Mech, GT2014bulk}. Other studies report dynamic response data that exhibit not only temperature- but also pressure-independent time-scale indices \cite{Roed2015PPE, Casa2016}.
	
	The observation of \textit{decoupling} in a set of response-function-specific time scales, i.e., the observation of a temperature-dependent time-scale index, is often attributed to the influence from other relaxation contributions rather than from the dynamics of the structural-relaxation process itself \cite{7liqs2005GT, ReinerZorn1997, GT2008monoalc}. In case of hydrogen-bonded liquids, which exhibit stronger intermolecular interactions than vdW-bonded liquids, the formation of supramolecular structures is presumably reflected by a slow relaxation process which can also affect the overall peak position \cite{C7CP06482A,gabriel2020GLY} and thus the coupling behavior.
	
	While the aforementioned studies mainly reported on the coupling tendencies that were observed for different sets of response functions, Jakobsen \textit{et al.} \cite{vdw2012GT} put a focus on the ordering of the time-scale indices connected to the probed response functions. There, time scales were determined for seven different response functions for the two van-der-Waals (vdW) bonded glass formers 5PPE and DC704. These two materials exhibit simple behavior in terms of a non-detectable secondary relaxation process and obedience to time-temperature superposition \cite{THKN2018}. Time scales differ significantly between response functions while their temperature-dependence is the same, and obey a response-function-specific ordering that is identical for both investigated liquids,	
	\begin{equation*}\label{eq:order}
		\tau_{G} < \tau_{K} < \tau_{\epsilon} < \tau_{\alpha_l} < \tau_{c_l}.
	\end{equation*}
	The time scales are based on measurements of the following response functions: the shear-mechanical modulus $G(\omega)$, the bulk-mechanical modulus $K(\omega)$, the dielectric permittivity $\epsilon(\omega)$, the longitudinal thermal expansion coefficient $\alpha_l(t)$, and the longitudinal specific heat $c_l(\omega)$ \cite{vdw2012GT}.
	
	To test how general the response-function-dependent ordering of time-scale indices is, more complex liquids have to be put under investigation. Two aspects that are relevant in this context are 1) the interference of additional relaxation processes with the primary relaxation process as it has an impact on time scales derived from spectral data, and 2) the influence of inter-molecular interactions on the different response functions.
	
	Thus, in this study, the dynamics of two non-simple liquids with significantly different properties regarding bonding type and relaxation processes are analyzed for five different experimental response functions. These two liquids are 1,2,6-hexanetriol (hexanetriol), a hydrogen-bonded liquid with a weak secondary contribution, and  2,6,10,15,19,23-hexamethyltetracosane (squalane), a vdW-bonded liquid with a strong secondary relaxation. Time-scale coupling, i.e., the observation of a temperature-independent time-scale index, is reported for hexanetriol, while squalane shows decoupling of the time scales if these are determined from spectral loss-peak positions. If time scales are derived from terminal modes of spectra measured on squalane, response-specific time scales exhibit less decoupling and identical response-specific ordering of time-scale indices is observed for both vdW-bonded and the hydrogen-bonded liquid.
	
	In the following, Sec. \ref{ExpMeth} describes briefly the methods used for obtaining the five different response functions. In Sec. \ref{Res}, data that were measured on these response functions for hexanetriol and squalane are subdivided into three parts. Sec.~\ref{sec:spectra} contains spectral data, sec.~\ref{sec:time scales} presents time scales of structural relaxation, and in Sec.~\ref{sec:index}, the time-scale indices obtained for hexanetriol and squalane are discussed and compared to results for the simple vdW-liquid DC704 as presented in Ref.~\onlinecite{vdw2012GT}.

	\section[Experimental]{Experimental methods}\label{ExpMeth}
	
	Temperature-dependent time-scale data were obtained for the following five response functions: the shear modulus $G(\omega)$, the bulk modulus $K(\omega)$, the dielectric permittivity $\epsilon(\omega)$, the longitudinal thermal expansion $\alpha_l(t)$, and longitudinal specific heat $c_l(\omega)$. The transducers used for measuring these response functions are customized for usage within the same cryostat (available temperature range from \SI{100}{\kelvin} to \SI{310}{\kelvin}), ensuring comparable thermal conditions involving steady absolute temperature over weeks and restricting temperature fluctuations to few \si{\milli\kelvin} \cite{GT2008cryo}. Response functions are determined with the same electronics system on the basis of electrical measurements \cite{GT2008imped}. Shear-mechanical measurements were conducted with the piezo-electric shear-modulus gauge, that allows for measurements at frequencies $\nu = \omega/(2\pi)$ between \SI{1}{\milli\hertz} and \SI{50}{\kilo\hertz} \cite{GT1995shear}. Bulk-modulus data were obtained by a piezo-electric bulk-modulus gauge (frequency range \SI{1}{\milli\hertz} to \SI{10}{\kilo\hertz}) \cite{GT1994bulk}. Dielectric permittivity data are based on measurements with a parallel-plate capacitor (fixed spacing of \SI{50}{\micro\meter}) within the available frequency range (\SI{1}{\milli\hertz} to \SI{1}{\mega\hertz}). For squalane, the dielectric data set was supplemented by measurements with an Andeen Hagerling 2700A high-resolution capacitance bridge (frequency range \SI{50}{\hertz} to \SI{20}{\kilo\hertz}).  Time-dependent thermal expansion data are based on high-resolution dielectric dilatometry measurements at a fixed frequency under fast temperature regulation using a parallel-plate capacitor in which the sample material controls the spacing \cite{GT2012CTE}. Dynamic specific-heat data were measured by a negative temperature-coefficient thermistor simultaneously serving as temperature gauge and heating device in the available frequency range from ca. \SI{5}{\milli\hertz} to \SI{20}{\hertz} \cite{GT2008speche,GT2010speche}. Short reviews of the different measurement principles are given in the supplemental information (SI).
	
	Experimental data were acquired for the two different molecular glass formers hexanetriol and squalane. Hexanetriol is a hydrogen-bonding trihydric alcohol ($T_g \cong$ \SI{200}{\kelvin}) with a secondary process occurring in form of an excess wing \cite{GT1994HEX,HEXdiel} and similar properties to glycerol. The slightly higher glass transition temperature, however, makes hexanetriol a more suitable candidate for meeting the temperature limitation and the narrow frequency  window of the specific-heat measurement technique  \cite{HEXdiel}. The complex, frequency-dependent shear modulus $G(\omega)$, bulk modulus $K(\omega)$, dielectric permittivity $\epsilon(\omega)$, and longitudinal specific heat $c_l(\omega)$ were measured for hexanetriol under identical experimental conditions, i.e., within the same cryostat and with the same electronic setup. As hexanetriol has a significant dipole moment, the determination of the thermal expansion coefficient based on dielectic dilatometry is not feasible \cite{GT2012CTE}. Hexanetriol was purchased from Sigma-Aldrich with a purity of \SI{96}{\percent} and used as received. The container was opened in a dry  N$_2$ atmosphere to avoid contamination with water. The sample cells were prepared and filled in a dry N$_2$ atmosphere, with the exemption of the bulk-modulus measurements, for which the transducer was filled in ambient air.
	
	Squalane is a vdW-bonded liquid ($T_g \cong$ \SI{168}{\kelvin}) with a strong secondary relaxation peak in both mechanical and dielectric data. The complex dynamic shear modulus $G(\omega)$, bulk modulus $K(\omega)$, and dielectric permittivity $\epsilon(\omega)$ measured in one cryostat are complemented by data of the longitudinal thermal expansion coefficient $\alpha_l(t)$ that was obtained in another cryostat. The temperatures of the two cryostats were calibrated based on peak positions of dielectric data, which were available for both cryostats, resulting in a shift of the expansivity data by \SI{0.80}{\kelvin}. Note that the bulk-modulus data set was published in Ref. \onlinecite{GaT2017FitABSQLN}, and the dynamic thermal-expansion data is from the experiment presented in Ref. \onlinecite{GT2012CTE}. All data sets are based on material from the same bottle.

	\section{Results}\label{Res}

	\subsection{Spectral data}\label{sec:spectra}
	In this section the frequency-dependent response functions for hexanetriol and squalane are presented. These are the real and imaginary parts of the shear modulus $G(\omega)$, the bulk modulus $K(\omega)$, the dielectric permittivity $\epsilon(\omega)$, and the longitudinal specific heat $c_l(\omega)$ in case of hexanetriol. For squalane, the loss contributions of the shear modulus $G''(\omega)$, the bulk modulus $K''(\omega)$, and the dielectric permittivity  $\epsilon''(\omega)$, are presented.

	\subsubsection{Hexanetriol}\label{SpectraHEX}
	
	\begin{figure}[t!]
		\centering
		\includegraphics[width=\columnwidth]{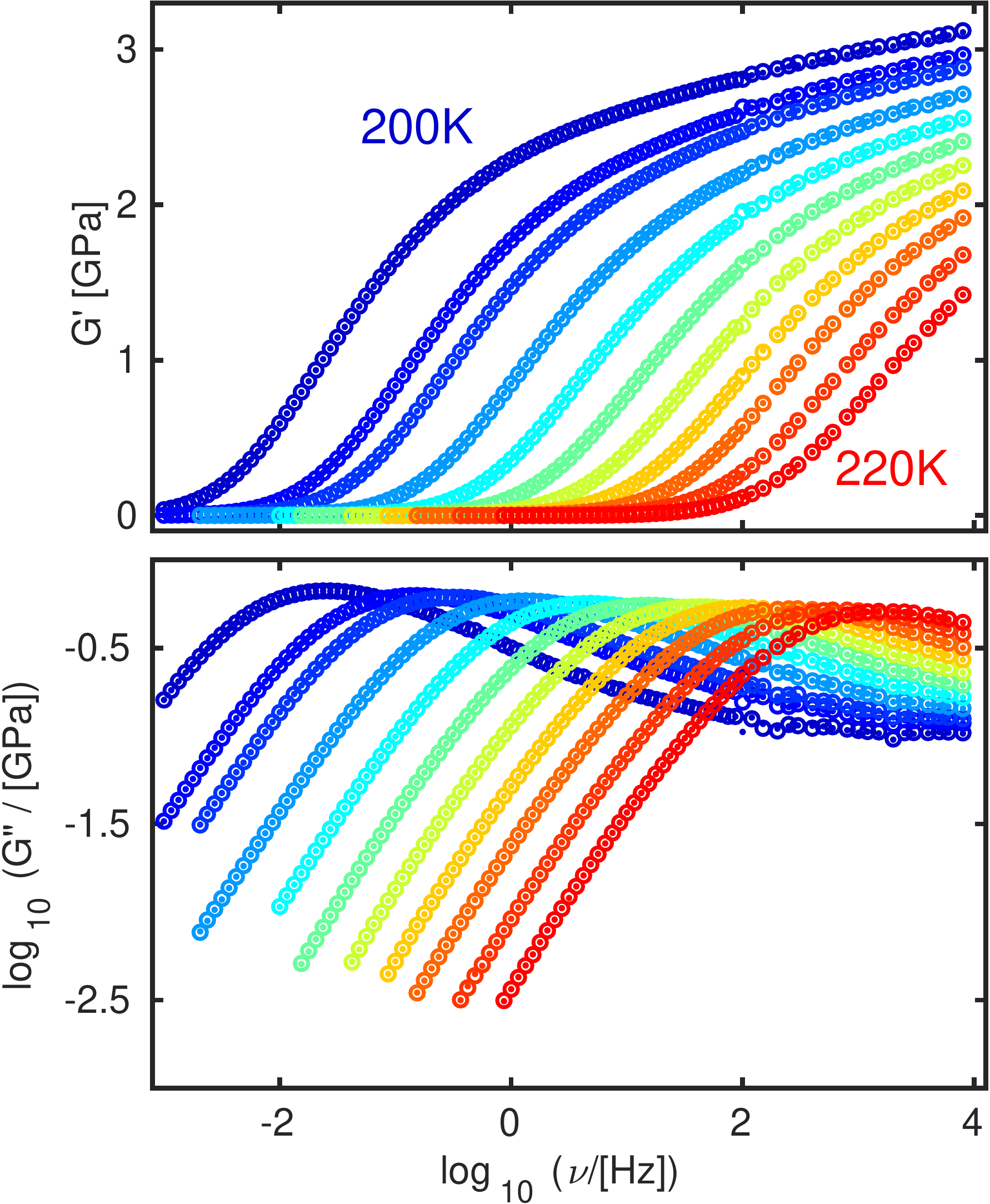}
		\caption{Storage and loss spectra of the dynamic shear modulus $G (\omega) = G' (\omega) + i G'' (\omega)$ of the supercooled liquid state of hexanetriol at selected temperatures ranging from \SI{200}{\kelvin} to \SI{220}{\kelvin}. After quenching to \SI{200}{\kelvin}, the shear modulus was measured in steps of \SI{1}{\kelvin} from \SI{202}{\kelvin} to \SI{214}{\kelvin} and steps of \SI{2.5}{\kelvin} from \SI{215}{\kelvin} to \SI{220}{\kelvin}. Points correspond to the first and open circles to the second, subsequently measured spectrum.}
		\label{FigHEXshear}
	\end{figure}
	
	\begin{figure}[t!]
		\centering
		\includegraphics[width=\columnwidth]{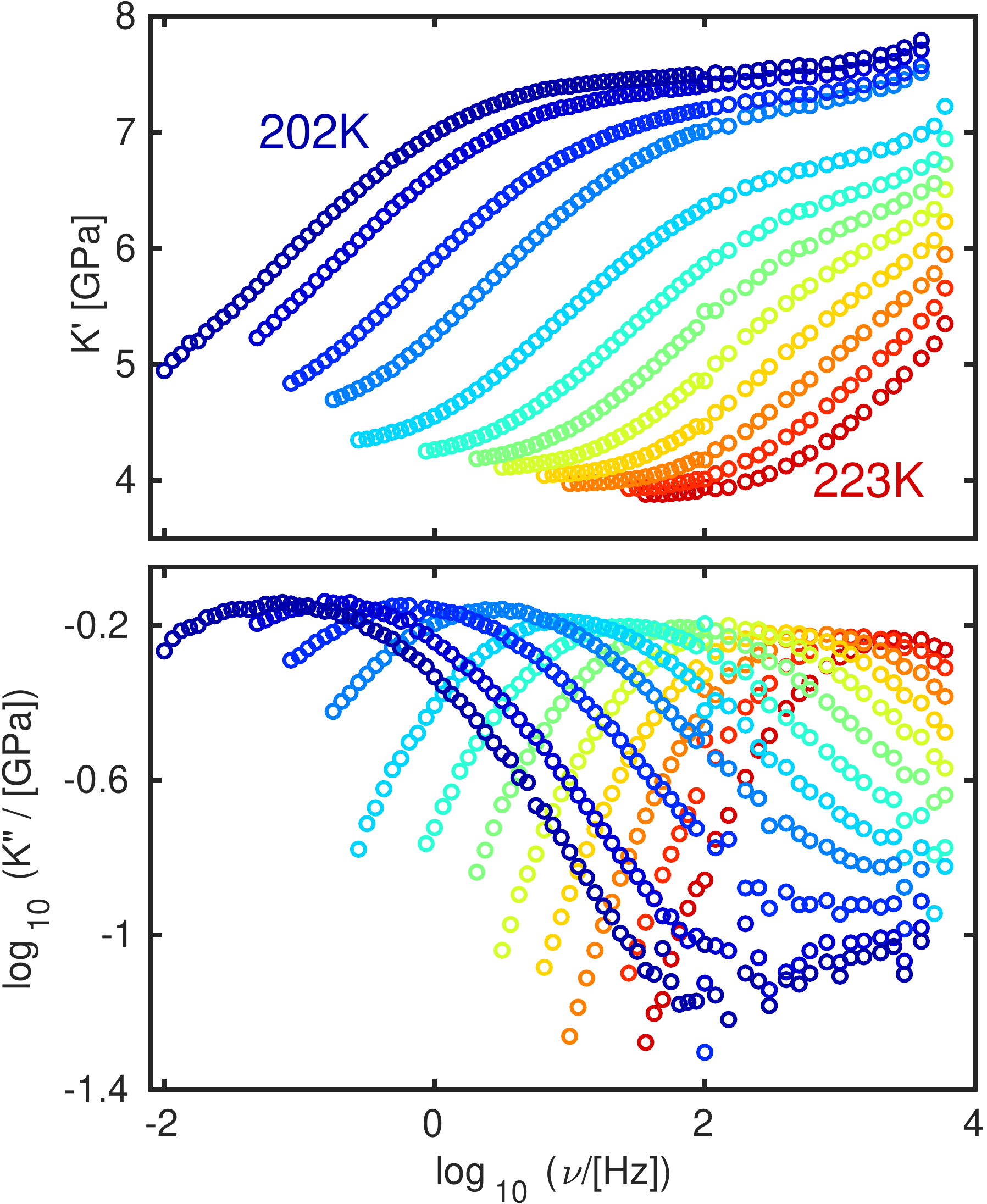}
		\caption{Storage and loss spectra of the dynamic bulk modulus $K (\omega) = K' (\omega) + i K'' (\omega)$ of the supercooled liquid state of hexanetriol for selected temperatures between \SI{202}{\kelvin} and \SI{223}{\kelvin}. Spectra were measured from \SI{223}{\kelvin} to \SI{203}{\kelvin} in steps of \SI{-2}{\kelvin} and from \SI{202}{\kelvin} to \SI{222}{\kelvin} in steps of \SI{+2}{\kelvin}.}
		\label{FigHEXbulk}
	\end{figure}

	The storage and loss contribution of the dynamic shear modulus are plotted in Fig.~\ref{FigHEXshear} for selected temperatures between \SI{200}{\kelvin} and \SI{220}{\kelvin}. At each temperature, two spectra were measured successively. As these data coincide within the accuracy of the measurement technique, full equilibration of the sample material at all depicted temperatures is assumed. Measurements of the dynamic bulk-mechanical modulus were conducted in the temperature range between \SI{202}{\kelvin} and \SI{224}{\kelvin} measured in \SI{1}{\kelvin} steps with a single spectrum measured at each temperature (Fig. ~\ref{FigHEXbulk}).
	
	Both the shear- and the bulk-modulus data show the charateristic features of a structural relaxation process. The storage contribution of the shear modulus approaches zero for frequencies $\nu \ll \nu_{lp}$, while the bulk modulus approaches the static contribution $K(0)$ in this limit. For both moduli the loss contributions exhibit a power-law behavior with exponent \num{+1} on the low-frequency flank. The spectra are rather broad, indicative of a slow relaxation process as previously reported for glycerol \cite{C7CP06482A,gabriel2020GLY}. Based on the conclusions drawn for glycerol, the relaxation feature of hexanetriol may derive from supramolecular structure dynamics reflecting interactions between the various OH-groups of adjacent molecules. Additionally, at all temperatures the signature of a secondary process is observed at high-frequencies, where both $G''(\omega)$ and $K''(\omega)$ exhibit a shoulder. This shoulder is more separated from the main process at low temperatures, but merges with the primary relaxation process at higher temperatures. This aspect is also evident in the high-frequency regime of the storage contribution, where the data steadily increase instead of reaching a constant plateau.\\

	\begin{figure}[t!]
		\centering
		\includegraphics[width=\columnwidth]{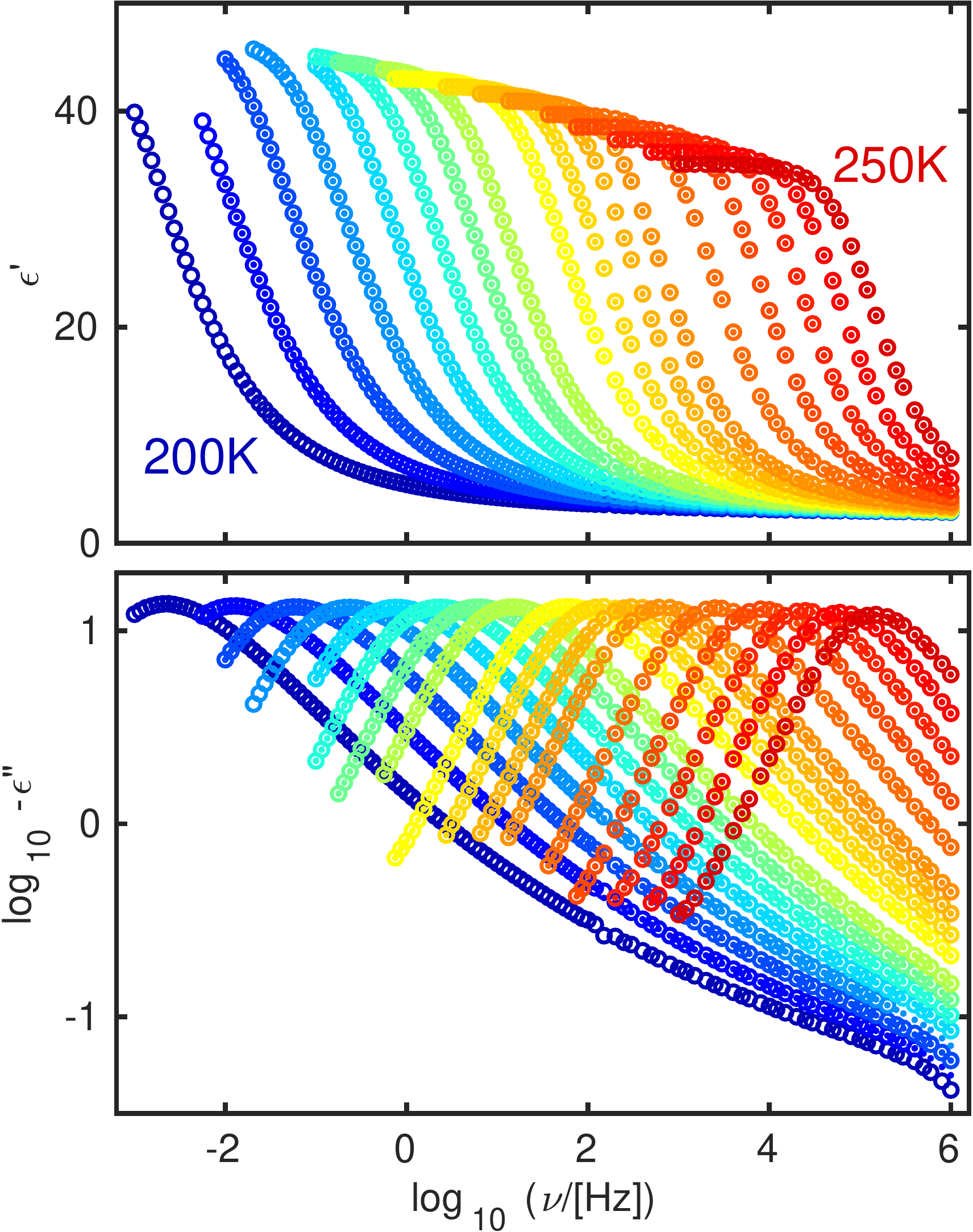}
		\caption{Storage and loss spectra of the dynamic dielectric permittivity $\varepsilon (\omega) = \varepsilon' (\omega) - i \varepsilon'' (\omega)$ of the supercooled liquid state of hexanetriol at selected temperatures ranging from \SI{200}{\kelvin} to \SI{250}{\kelvin}. After a quench to \SI{200}{\kelvin}, spectra were measured in temperature steps of \SI{1}{\kelvin} up to \SI{215}{\kelvin}, steps of \SI{2.5}{\kelvin} from \SI{217.5}{\kelvin} to \SI{225}{\kelvin}, and steps of \SI{5}{\kelvin} from \SI{230}{\kelvin} to \SI{250}{\kelvin}. Points correspond to the first, open circles to the second, subsequently measured spectrum.}
		\label{FigHEXdiel}
	\end{figure}
	
	The dielectric permittivity was measured at 25 temperatures between \SI{200}{\kelvin} and \SI{250}{\kelvin}. The loss and storage contributions are plotted in Fig.~\ref{FigHEXdiel} for selected temperatures. At most temperatures, two subsequent spectra were measured and the collapse of these demonstrate a well-equilibrated state of the material. The dielectric loss spectra show the expected power law on the low-frequency flank with an exponent of \num{1}, as well as powerlaw behavior with an exponent of \num{-1/2} on the high-frequency flank, indicating that the influence of the secondary relaxation contribution is negligible within the measured range of dielectric data. As is the case for glycerol, hexanetriol is expected to form a branched network of OH-bonds that gives no directional preference for an effective dipole moment \cite{C7CP06482A}, reflected by the absence of a slow process on the low-frequency flank.\\

	\begin{figure}[t!]
		\centering
		\includegraphics[width=\columnwidth]{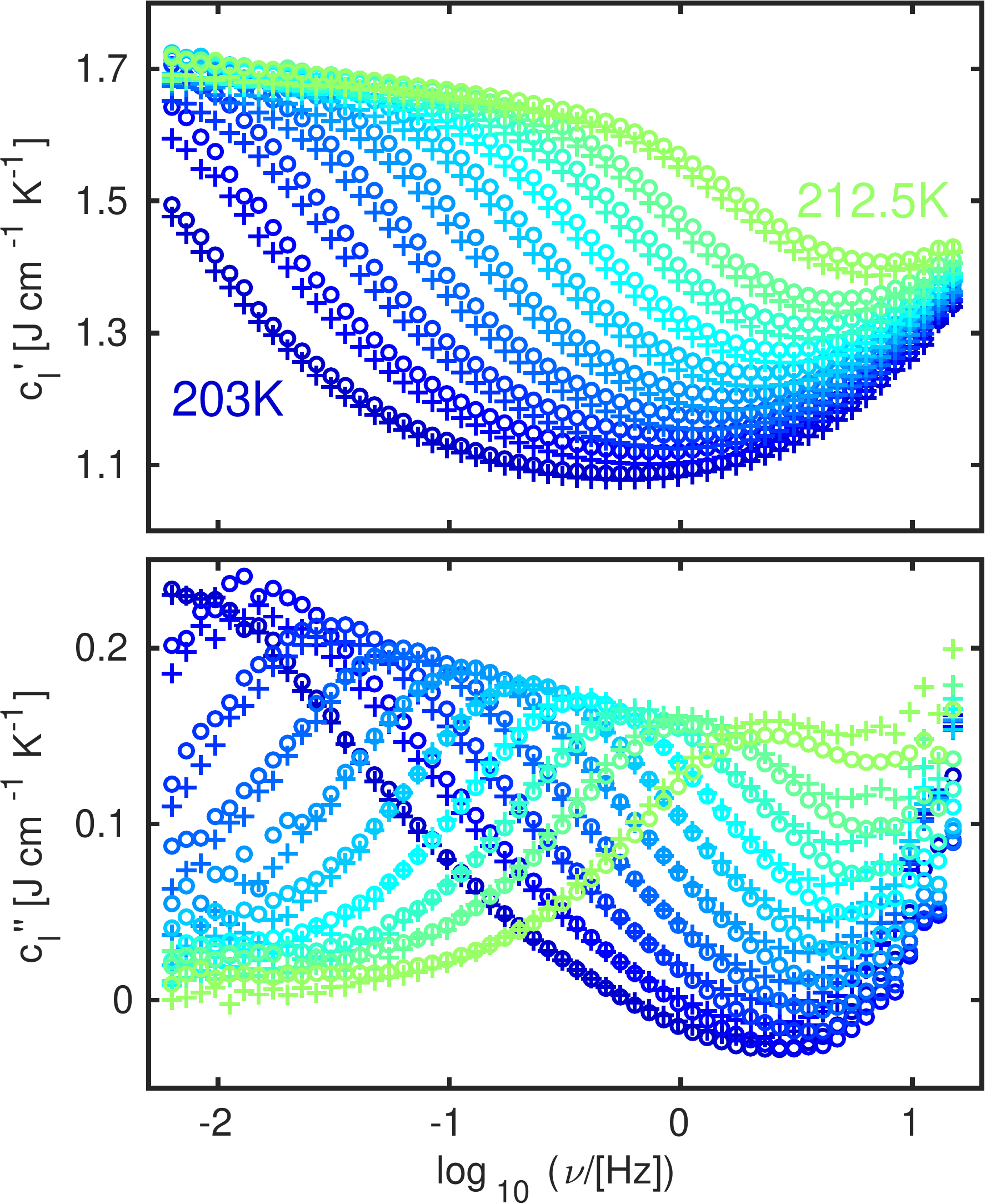}
		\caption{Storage and loss spectra of the dynamic longitudinal heat capacity $c_l (\omega)$ in the supercooled liquid state of hexanetriol at temperatures ranging from \SI{203}{\kelvin} to \SI{212.5}{\kelvin}. Crosses and circles reflect measurements at two different temperature amplitudes.}
		\label{FigHEXheat}
	\end{figure}

	Figure~\ref{FigHEXheat} shows loss and storage contribution of the dynamic longitudinal heat capacity $c_l (\omega)$ at ten temperatures in the supercooled liquid, measured at two different temperature amplitudes. Both amplitudes yield a response within the linear regime and deviations between spectra at a given temperature indicate the uncertainty of the data. Two distinct artifacts are observed for $c_l (\omega)$, emphasizing that the accuracy of the details of the measured spectra of longitudinal heat capacity is limited compared to the other measured response functions: First, spectra measured at low temperatures exhibit slightly negative values of the imaginary part in the frequency range around \SI{1}{\hertz}. Second,	both storage and loss contributions show increasing values at high frequencies. If the increase in $c_l''$ were solely due to a secondary contribution, $c_l'$ would show a stepwise decrease towards a high-frequency plateau based on Kramers-Kronig relations. Despite these issues, normalizing the loss spectra to the amplitude and position of the maximum results in a reasonable collapse (see Fig. S1d in SI). This emphasizes that the spectra obey time-temperature-superposition despite the limited accuracy in details of their shape. Thus it seems reasonable to extract $\nu_{lp}$ for comparison to other time scales.	
	
	\subsubsection{Squalane}\label{SpectraSQLN}

	\begin{figure}[t!]
		\centering
		\includegraphics[width=\columnwidth]{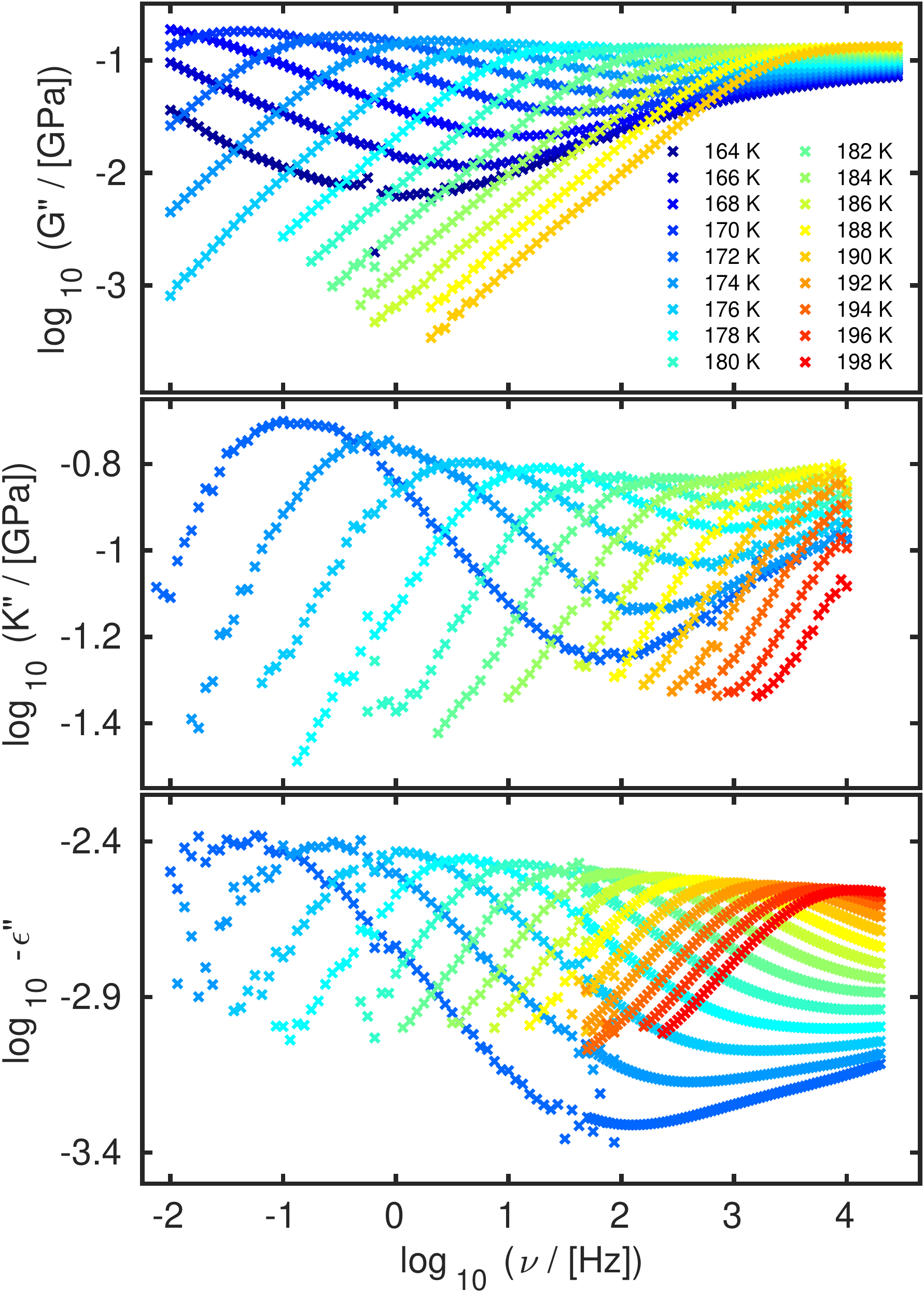}
		\caption{Loss spectra of the dynamic shear modulus $G (\omega)$, bulk modulus $K (\omega)$, and dielectric permittivity $\varepsilon(\omega)$ of squalane from \SI{164}{\kelvin} to \SI{190}{\kelvin} for shear-mechanical data, and from \SI{172}{\kelvin} to \SI{198}{\kelvin} for dielectric and bulk-mechanical data.}
		\label{FigSQLNloss}
	\end{figure}
	For a range of temperatures, Fig.~\ref{FigSQLNloss} shows the loss contributions $\epsilon''(\omega)$, $G''(\omega)$, and $K''(\omega)$ of squalane as functions of frequency. All three response functions show qualitatively comparable features: A primary relaxation contribution with a strongly temperature-dependent loss-peak position and a secondary contribution at higher frequencies with a comparably static loss-peak position, which merge upon increasing temperature.
	
	However, details of the loss spectra of these three response functions differ significantly in their low-frequency behavior. The shear-mechanical data exhibit a power law behavior with an exponent close to unity (see fig.~S1 in the SI), which thus resembles the rather universal behavior reflected by the response functions measured on hexanetriol. In case of the bulk-mechanical and dielectric responses of squalane, however, the exponent of the low-frequency flank is significantly reduced to values around \num{0.3} to \num{0.4}. This behavior might be a consequence of the rather small signal amplitudes of the two data sets, especially in the case of the dielectric data. 	
	
	\subsection{Time scales of structural relaxation}\label{sec:time scales}
	
		\begin{figure*}[t]	
		\begin{minipage}{\columnwidth}
			\centering
			\includegraphics[width=\textwidth]{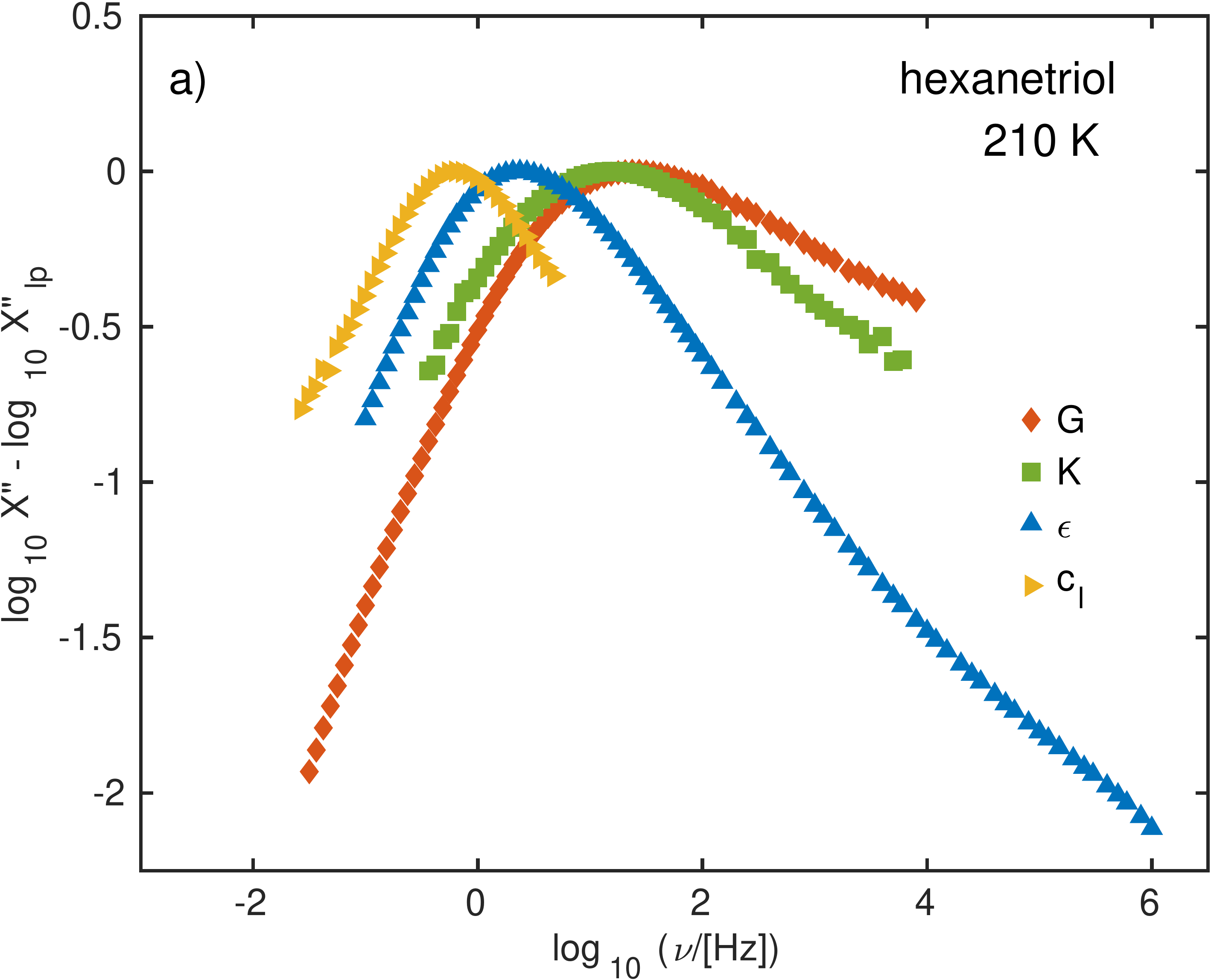}
		\end{minipage}
		\begin{minipage}{\columnwidth}
			\centering
			\includegraphics[width=\textwidth]{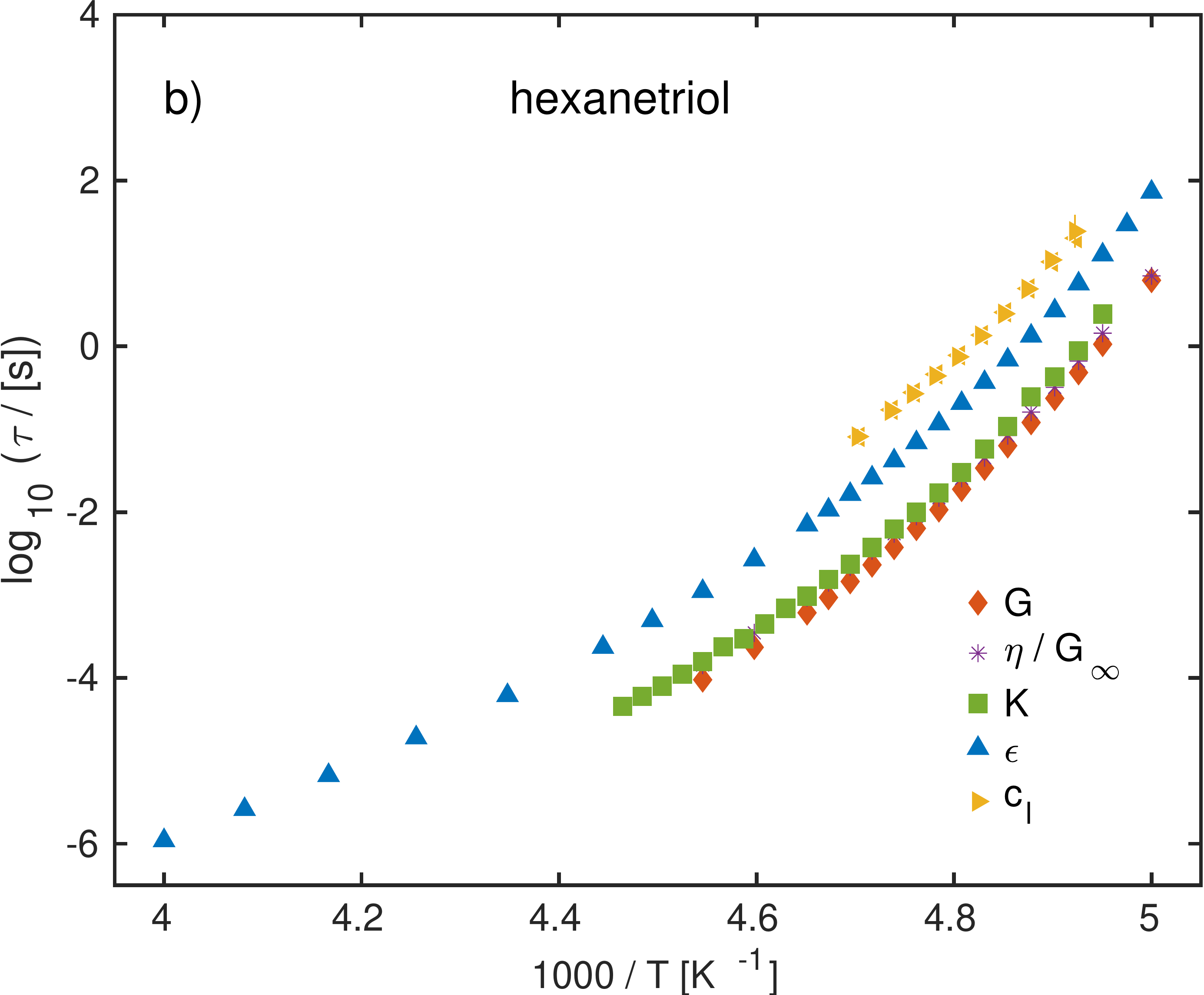}
		\end{minipage}	
		\caption{\textbf{(a)} Loss contributions of four complex frequency-dependent response functions: the shear modulus $G(\omega)$, the bulk modulus $K(\omega)$, the dielectric permittivity $\epsilon(\omega)$, and the longitudinal specific heat $c_l(\omega)$, as functions of frequency at \SI{210}{\kelvin} for hexanetriol, normalized to the individual loss-peak height. \textbf{(b)} time scales based on spectral loss-peak positions of the four response functions depicted in (a) and Maxwell time scale as functions of inverse temperature for hexanetriol.}
		\label{FigHEX}
	\end{figure*}
	\begin{figure*}[t]	
		\begin{minipage}{\columnwidth}
			\centering
			\includegraphics[width=\textwidth]{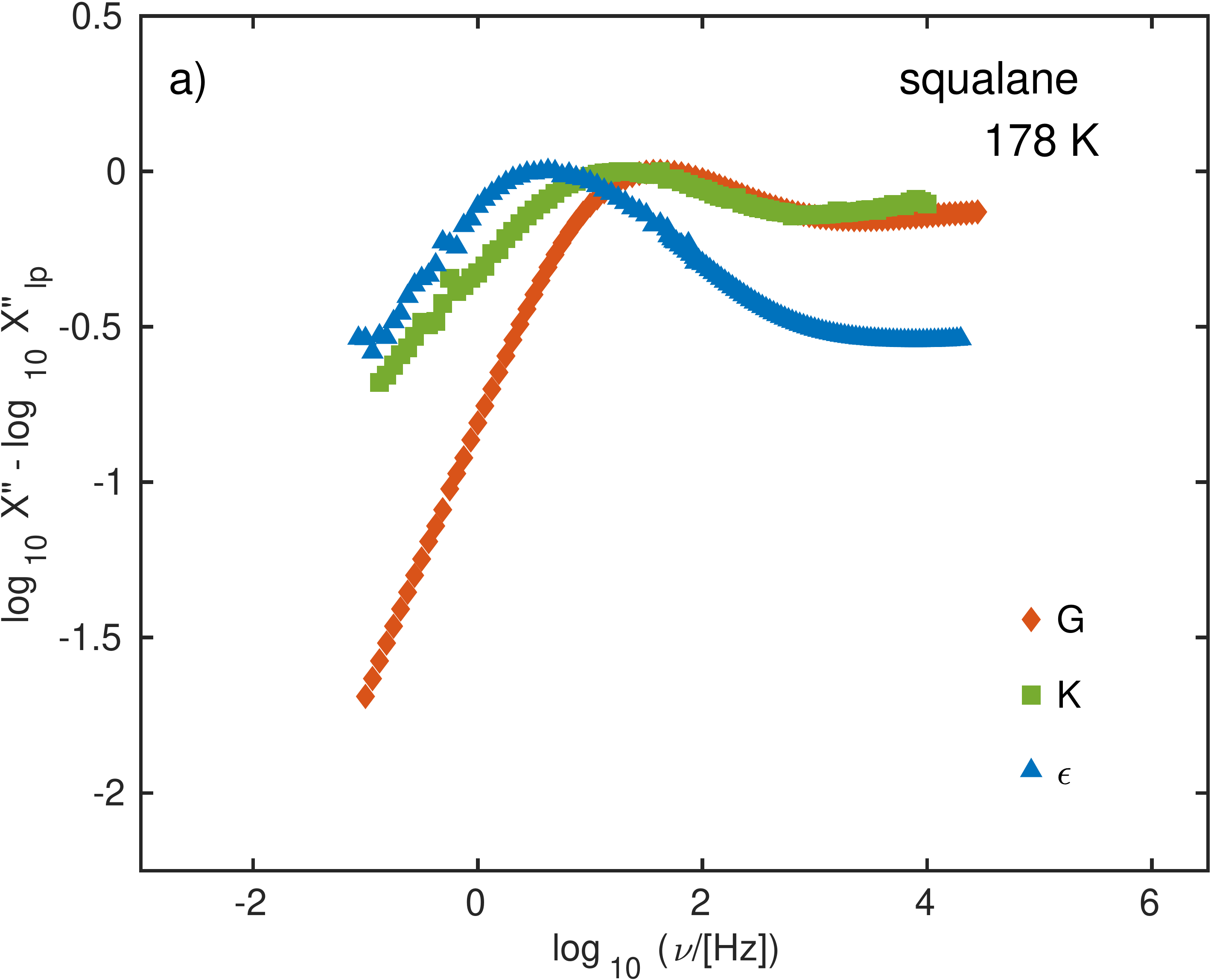}
		\end{minipage}
		\begin{minipage}{\columnwidth}
			\centering
			\includegraphics[width=\textwidth]{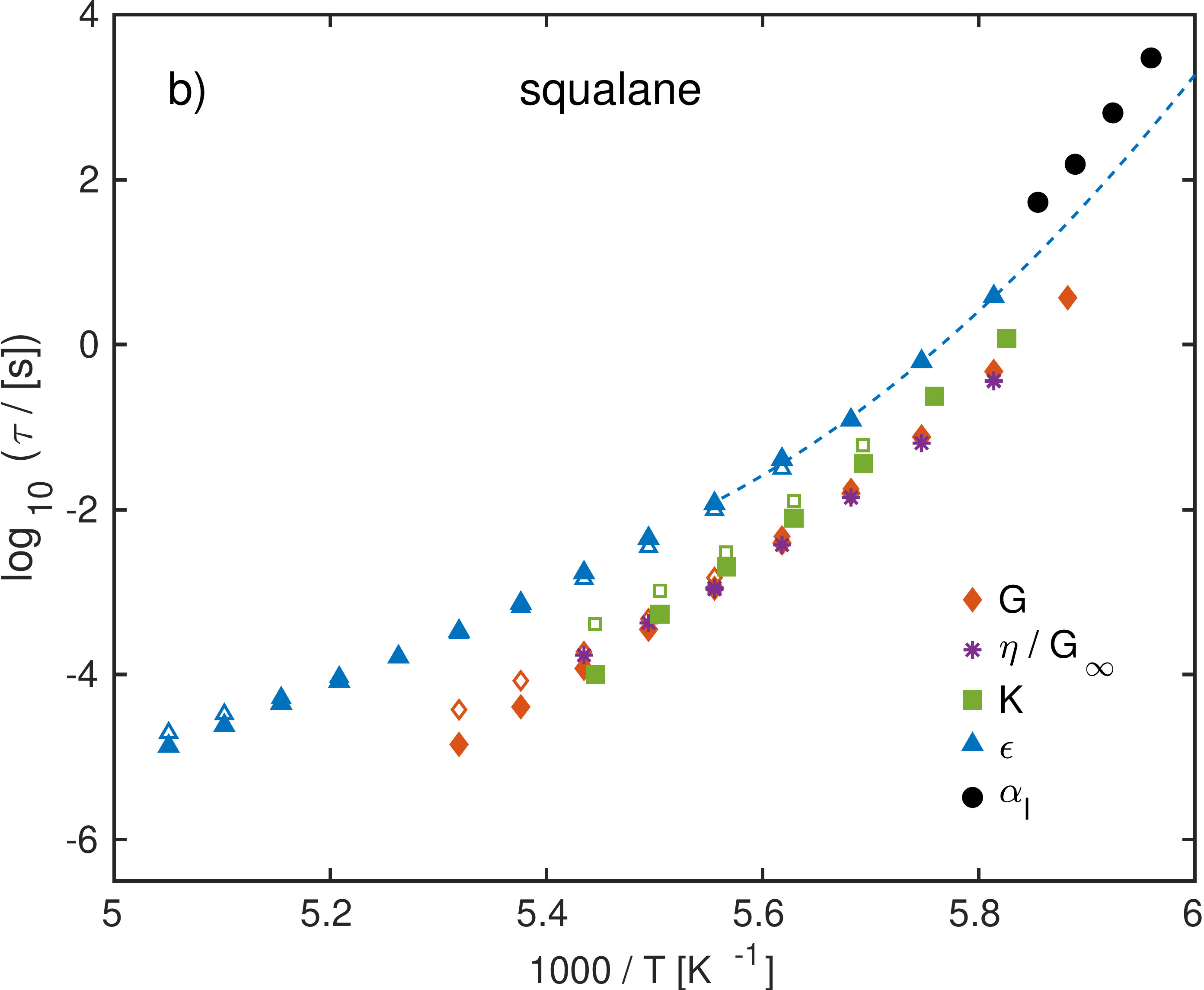}
		\end{minipage}	
		\caption{\textbf{(a)} Loss contributions of three complex frequency-dependent response functions: the shear modulus $G(\omega)$, the bulk modulus $K(\omega)$, and the dielectric permittivity $\epsilon(\omega)$, as functions of frequency at \SI{178}{\kelvin} for squalane, normalized to the individual loss peak height. \textbf{(b)} Filled symbols: time scales based on spectral loss-peak positions of the three response functions depicted in (a) and time scales derived from expansivity data, plotted as functions of the inverse temperature for squalane. Open symbols: time scales based on the spectral shift of the low-frequency flank relative to a low-temperature loss-peak-position derived time scale. The dashed line corresponds to a fit of the dielectric data between \SI{172}{\kelvin} and \SI{180}{\kelvin} to a parabolic function with the parameters $\log_{10} \tau_0 = $\num{-2.96}, $J = $\SI{3308}{\kelvin}, and $T_0 = $\SI{190.6}{\kelvin}.}
		\label{FigSQLN}
	\end{figure*}

	\begin{figure*}[t!]	
		\begin{minipage}{\columnwidth}
			\centering
			\includegraphics[width=\textwidth]{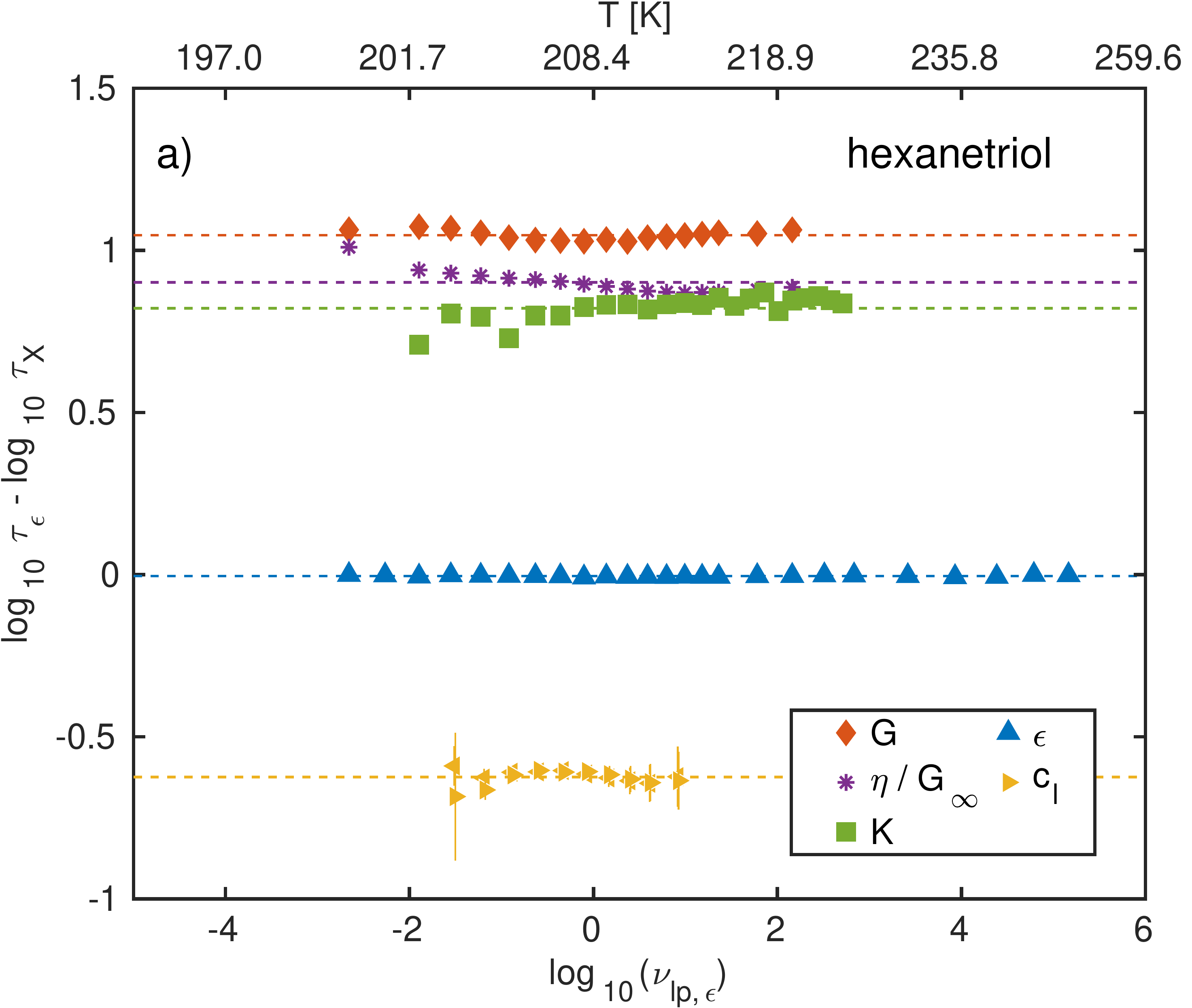}
		\end{minipage}	
		\begin{minipage}{\columnwidth}
			\centering
			\includegraphics[width=\textwidth]{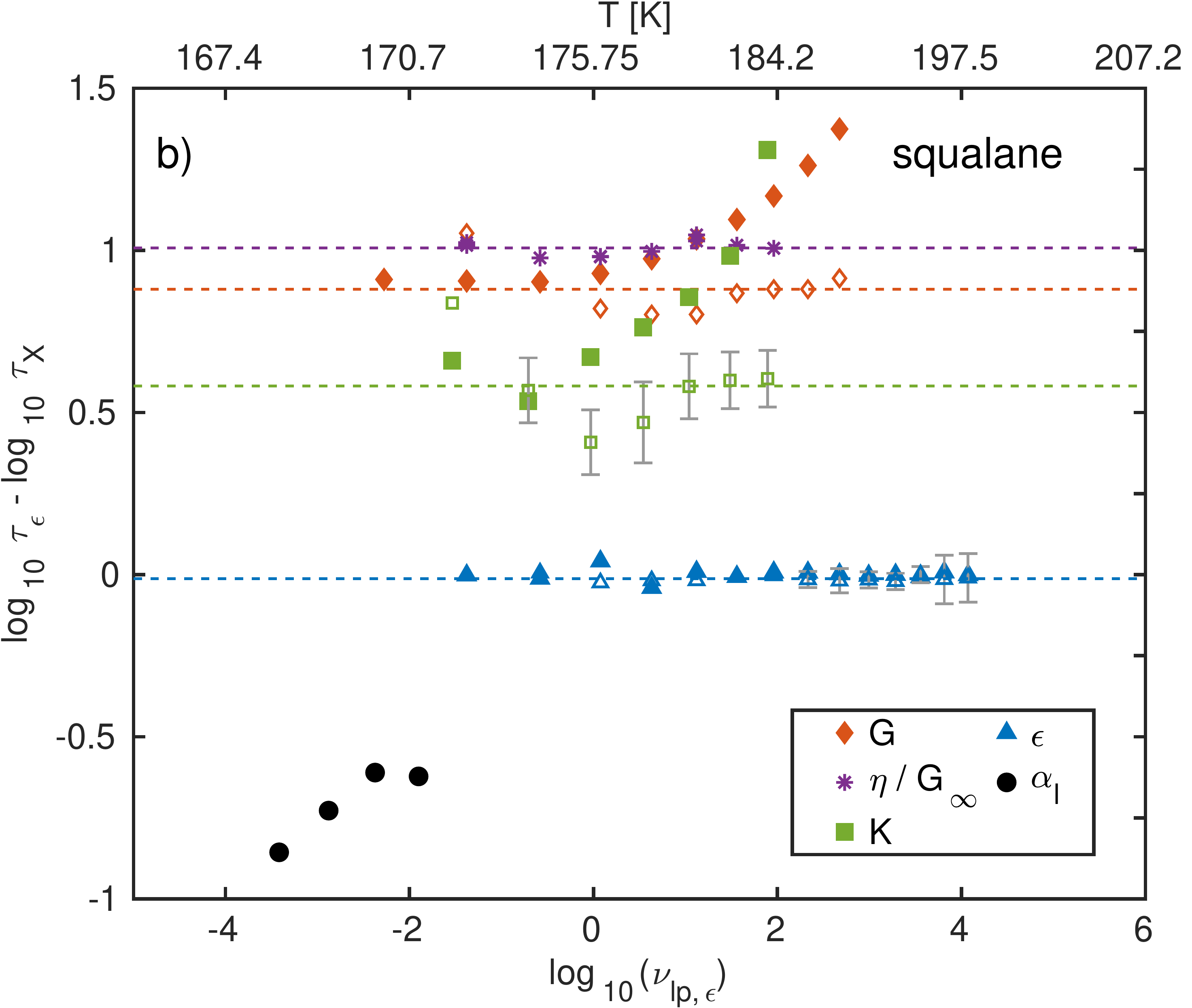}
		\end{minipage}
		\caption{Time-scale index for several response functions based on spectral loss-peak position (filled symbols) and terminal modes (open symbols) for hexanetriol in \textbf{(a)} and squalane in \textbf{(b)} plotted as functions of the dielectric loss-peak frequency (bottom x-axis) and the temperature (top x-axis). Dashed lines represent average values of the time-scale indices, based on time scales derived from the loss-peak maximum for hexanetriol and the terminal mode for squalane.}
		\label{FigIndex}
	\end{figure*}

	Loss spectra of the four response functions $G(\omega)$, $K(\omega)$, $\epsilon(\omega)$, and $c_l(\omega)$ recorded at \SI{210}{\kelvin} are plotted as functions of frequency for hexanetriol in Fig.~\ref{FigHEX}(a). In this direct comparison it can be seen that the specific heat and dielectric loss-peaks are more narrow than those of the mechanical data. Both bulk- and shear-mechanical moduli show rather rounded features on the low-frequency flank, indicating the existence of an additional, slow relaxation process as reported for glycerol \cite{C7CP06482A,gabriel2020GLY}.
	
	The maximum-loss frequency $\nu^{lp}_X$ was identified as the loss-peak frequency for each frequency-dependent response function $X$. Assuming that the slow process proposed for hexanetriol has a negligible influence on the peak position, and as the spectra only show a minor secondary process, $\tau_X = ( 2 \pi \nu^{lp}_X )^{-1}$ reflects the time scale of structural relaxation. From $G(\omega)$, the Maxwell time scale $\tau_M$ was determined based on the relation $\tau_M = \eta_0 / G_{\infty}$ in which the viscosity $ \eta_0  = G''(\omega)/\omega |_{\omega \rightarrow 0}$ was deduced from the mechanical data and $G_{\infty}$ was found from fits of the shear-mechanical data to an electrical-equivalent-circuit (EEC) model (see SI for details on fitting function and parameters). The temperature dependence of the five response-function-specific time scales for hexanetriol is depicted in Fig.~\ref{FigHEX}(b). Note that the order of these time scales at a given temperature, as shown in Fig.~\ref{FigHEX}(a), is maintained throughout the temperature range over which all response functions were measured (\SI{203}{\kelvin} to \SI{212.5}{\kelvin}), covering changes in $\tau$ of more than two orders of magnitude.  The time scales determined from the two mechanical measurements are very similar, with shear-mechanical data showing slightly faster response behavior than the dynamic bulk-mechanical data. The dielectric and specific heat data exhibit a significantly slower relaxational response at a given temperature compared to the other response functions. Thus, hexanetriol exhibits the same ordering as simple vdW-bonded liquids \cite{vdw2012GT}.
	
	For squalane, the loss spectra of $G(\omega)$, $K(\omega)$, and $\epsilon(\omega)$, measured at \SI{178}{\kelvin} are plotted in Fig.~\ref{FigSQLN}(a). For all three response functions, a secondary relaxation contribution is seen as an onset of a second peak on the high-frequency part of the spectra. In contrast to the shear-mechanical response, the low-frequency flanks of bulk-mechanical and dielectric data are broadened. These features might be connected to the small dielectric response amplitudes for squalane and a presumably non-ideal match of bulk-mechanical data with its reference measurement, emphasizing that the data acquisition with these two methods is pushed to its limits due to the low response amplitudes of squalane.	
	The squalane time scales extracted from the spectral positions of the loss peaks are plotted against temperature in Fig.~\ref{FigSQLN}(b). The time scales connected to the thermal expansion coefficient, $\tau_{\alpha_l}$, were determined from Laplace-transformed time-domain data measured by dielectric dilatometry. As for hexanetriol, the mechanical data show smaller time scales than the dielectric measurements, while the expansivity data exhibit the longest time scales. The latter conclusion is based on an extrapolation of the temperature dependence of dielectric data towards lower temperatures by a parabolic function $\tau(T) = \tau_0 \cdot \exp \left( J^2 (1/T - 1/T_0)^2 \right)$ (dashed line in Fig.~\ref{FigSQLN}(b)) \cite{vdw2012GT, FitParabol2010} as there is no overlap in temperature for dielectric and expansivity data.\\
	When comparing $\tau_{M}$ and the loss-peak-derived $\tau_{G}$, a deviation between the two is observed at high temperatures. A second approch was applied to extract the time scales of squalane, which focuses on the sprectral position of the terminal mode. To do so, spectra were manually shifted along the frequency-axis to collapse on the low-frequency flank. Relative to a loss-peak-derived time scale in the low-temperature regime of a data set, the time-scale indices were determined from the time-temperature-superposition related shift (open symbols in Fig.~\ref{FigIndex}(b)). The error bars indicate the range of uncertainty for the shift factor that results from the scatter of the spectral data.
	
	\subsection{Time-scale indices}\label{sec:index}
	
	To evaluate the temperature dependence of the time scales for the different response functions of hexanetriol and squalane, in the following we define the time-scale index as the logarithm of the ratio of the dielectric time scale and the time scale of another response function $X$, $\log \tau_{\epsilon} - \log \tau_X$. The normalization utilizes an interpolation of measured values of $\tau_{\epsilon}$ within the temperature range of dielectric data for each material. At temperatures below this range, an extrapolation by a parabolic function based on the low-temperature dielectric data is applied \cite{vdw2012GT, FitParabol2010}. Thus, in case of squalane, time-scale indices for expansivity and low-temperature shear-mechanical data are based on extrapolated dielectric dynamics and will depend on the function used for extrapolation. A parabolic function was applied as it resulted in the most robust extrapolation of experimental time-scale data in Ref.~\cite{vdw2012GT}. Deriving $\tau_{\alpha_l}$ from time-domain data and extrapolating the low-temperature dielectric time scale lead to a higher uncertainty of the time-scale index for expansivity data and do thus not allow for detailed conclusions on the coupling tendency.
	
\begin{figure}[t!]
		\includegraphics[width=\columnwidth]{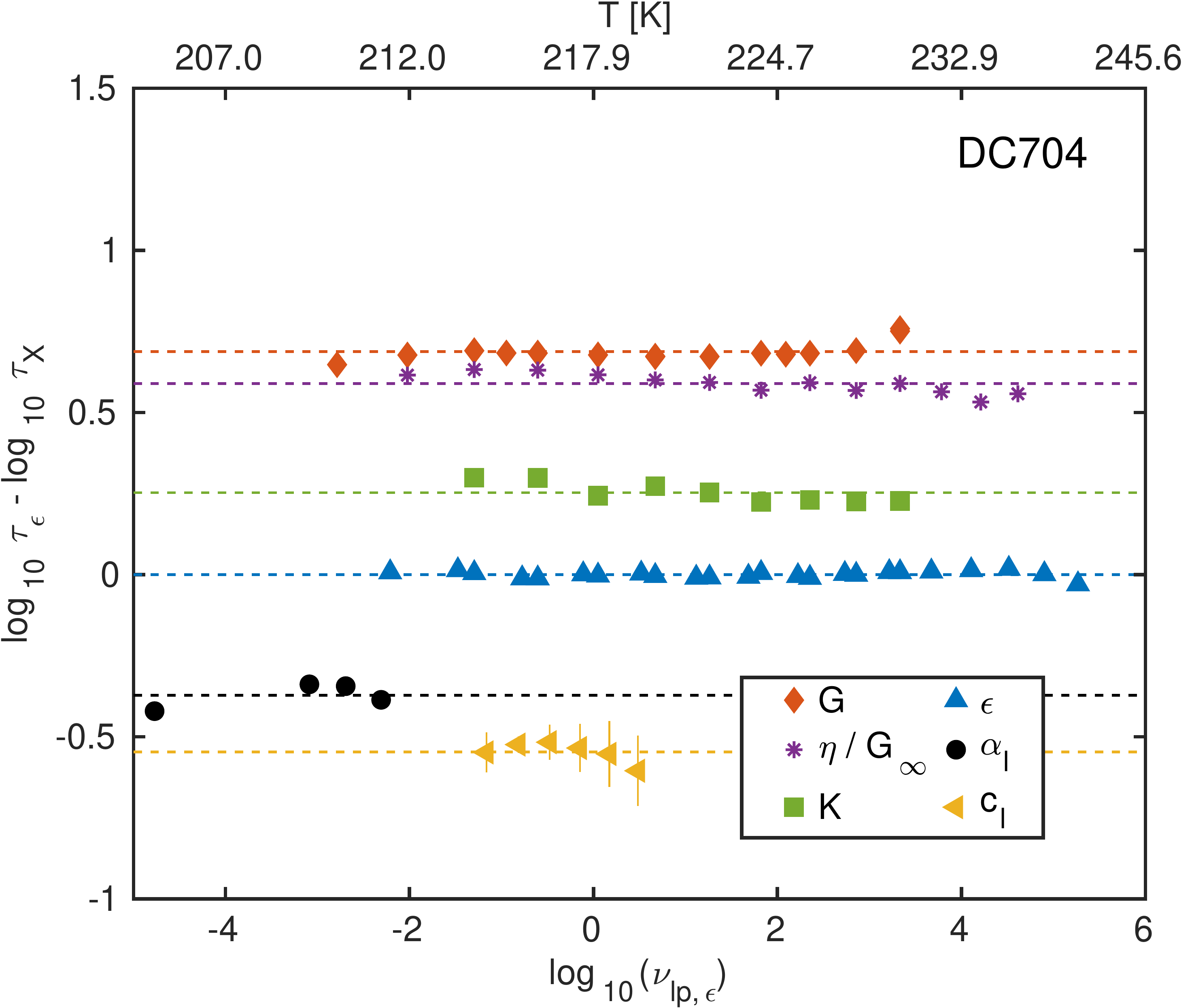}
		\caption{Time-scale index for several response functions for DC704 plotted as functions of the dielectric loss-peak frequency (bottom x-axis) and temperature (top x-axis) from Ref.~\onlinecite{vdw2012GT}.}
		\label{FigIndexdc}
	\end{figure}

	In Fig.~\ref{FigIndex}(a) the time-scales indices are plotted for hexanetriol. While $\tau_{G}$ is based on the spectral position of the loss peak maximum, $\tau_{M}$ reflects the position of the terminal mode at the low-frequency flank of the relaxation spectrum. Both time scales exhibit a temperature-independent time-scale index close to \num{1} for hexanetriol, suggesting only minor influence on the temperature dependence from the slow relaxation mode that is observed in the mechanical spectra \cite{C7CP06482A}. $\tau_G$ and $\tau_K$ are approximately one order of magnitude smaller than $\tau_{\epsilon}$, which is half a magnitude smaller than the time-scale index of the heat capacity. The error on $\tau_{c_l}$ is determined based on the spectral positions of the loss-peak maximum before and after correction for method-specific temperature-independent features of the specific-heat data (see Ref.~\onlinecite{GT2010speche} for details). From this set of response functions it becomes obvious that the specific time scales can differ by as much as 1.5 orders of magnitude at a given temperature.
	The general behavior of the time-scale indices for hexanetriol resembles that of DC704 (\ref{FigIndexdc}): For both materials the time-scale indices are temperature-independent and the ordering of response-specific time scales is identical, suggesting that the intermolecular interactions of vdW- and hydrogen-bonded materials affect the coupling of time scales in a similar fashion. These findings are in accordance with data reported for other alcohols \cite{Kremer2018, GT1994HEX, GT2008monoalc, C7CP06482A, HEXdiel}.
	
	In contrast to the behavior observed for hexanetriol and DC704, time scales derived from loss-peak positions for squalane lead to temperature-dependent indices for the mechanical and dielectric response functions, as shown in Fig.~\ref{FigIndex}(b). In previous studies, the apparent decoupling of time-scale indices for squalane was attributed to the influence of the secondary relaxation contribution on the position of the loss-peak maximum. Similar behavior was observed for other glass-forming liquids with a significant secondary relaxation contribution \cite{7liqs2005GT, ReinerZorn1997, GT2008monoalc}. $\tau_{M}$ however, related to the terminal relaxation mode, shows temperature-independent behavior. In analogy to $\tau_{M}$, and in order to diminish the effect of the secondary contribution, time scales for dielectric, shear-mechanical and bulk-mechanical data were extracted from their terminal mode. The time scales based on the terminal mode lead to less temperature-dependent time-scale indices for all tested response functions.This strengthens the view that the secondary relaxation contribution is the origin of the time-scale decoupling reported in literature for squalane. When considering the time-scale indices derived from the terminal mode, squalane clearly conforms to the same ordering of time scales as hexanetriol and DC704.
	
	\section[Conclusion]{Summary}
	
	This study presents time-scale data based on five different response functions with a focus on the coupling tendency and the ordering of time scales. It is based on two liquids, hexanetriol and squalane, representing hydrogen-bonded liquids with a minor secondary relaxation contribution and vdW-binded liquids with a prominent secondary relaxation process. These data are set into context to results on the simple vdW-bonded liquids from Ref.~\onlinecite{vdw2012GT}.
	
	The two major findings of the present study are: 1) The general observation of time-scale coupling if the structural relaxation process dominates the response behavior. 2) Identical ordering of response-function-specific time-scale indices. These observations are made for  all the aforementioned liquids despite their fundamentally different inter-molecular interactions (vdW-bonded or hydrogen-bonded liquids) and the varying degree of contributing relaxation processes (non-detectable, minor, strong secondary relaxation, or slow process) that occur in addition to the primary relaxation process.
	
	In more detail, the poly-alcohol hexanetriol exhibits coupling of the time scales of structural relaxation over the full investigated temperature range based on data from four response functions. This is in accordance with observations on other poly-alcohols with minor secondary contribution, and it strengthens the view that the type of inter-molecular bonding has negligible effect on the coupling tendency for the investigated liquids. 
	Continuing the discussion of the coupling behavior, squalane seems to oppose the observation made on hexanetriol on first glance, as it shows clear signs of decoupling for time scales determined from spectral loss peak positions. This observation has also been observed for other materials with a dominant secondary relaxation contribution \cite{7liqs2005GT, ReinerZorn1997}, however, this behavior can be attributed to the influence of the secondary relaxation contribution rather than the dynamics of structural relaxation itself. This is confirmed by time-scale indices determined from the dynamics of the terminal mode, which is much less affected by the secondary relaxation contribution and result in approximately temperature-independent, i.e. constant time-scale indices.
	
	Moreover, hexanetriol and squalane display the same order of response-function-specific time-scale indices as the simple vdW-bonded liquids from Ref.~\onlinecite{vdw2012GT}, indicating generic behavior for vdW- and hydrogen-bonded liquids:
	\begin{equation*}\label{eqo}
		\tau_{G} < \tau_{K} < \tau_{\epsilon} < \tau_{\alpha_l} < \tau_{c_l}
	\end{equation*}

	This finding, together with the observed time-scale coupling, indicates that the time scale ordering may be a more general feature of viscous liquids, extending beyond the simple liquids. It is especially surprising given that for the more complex liquids studied here, the spectral shapes of the different response functions are vastly different. For the simple liquids studied in Ref. \onlinecite{vdw2012GT}, the spectral shapes were nearly identical, and it was conjectured, that these features are connected to the concept of simplicity \cite{THKN2018}.
	As time-scale coupling and ordering hold for more complex vdW- and hydrogen-bonded liquids as well, it is reasonable to assume that the key characteristics of structural relaxation are of similar nature in all these types of liquids, and are explicitly not dominated by the type of inter-molecular bonding. However in complex materials the structural relaxation is often obscured by additional relaxation processes which lead to more complicated spectral shapes and differences in spectra obtained from different response functions. In agreement with this assumption, a comparison of spectral data based on light-scattering and dielectric spectroscopy for glycerol shows that self- and cross-correlated responses can be disentangled and strongly suggest comparable spectral shapes for the primary relaxation contributions measured by the two different response functions \cite{gabriel2020GLY}. This view is further supported by the fact that in many cases the relaxation spectra can be fitted by assuming several spectral contributions while shape parameters of the function describing the alpha relaxation are kept fixed \cite{C7CP06482A,Blochowicz2016}.	
	
	\section{Data Availability}
	
	The data that support the findings of this study are available at http://glass.ruc.dk/data/.
	
	\begin{acknowledgments}
	
	This work was funded by the VILLUM Foundation’s Matter Grant (No. 16515).
	
	\end{acknowledgments}

	\bibliographystyle{ieeetr}
	\bibliography{Man_TS}

\begin{thebibliography}{10}

\bibitem{angell1985strong}
C.~Angell, ``Strong and fragile liquids,'' {\em Relaxations in complex
  systems}, vol.~3, no.~11, 1985.

\bibitem{dyr06}
J.~C. Dyre, ``The glass transition and elastic models of glass-forming
  liquids,'' {\em Rev. Mod. Phys.}, vol.~78, pp.~953--972, 2006.

\bibitem{FitParabol2009}
Y.~S. Elmatad, D.~Chandler, and J.~P. Garrahan, ``Corresponding states of
  structural glass formers,'' {\em The Journal of Physical Chemistry B},
  vol.~113, no.~16, pp.~5563--5567, 2009.
\newblock PMID: 19254014.

\bibitem{pabst2017}
F.~Pabst, J.~Gabriel, P.~Weigl, and T.~Blochowicz, ``Molecular dynamics of
  supercooled ionic liquids studied by light scattering and dielectric
  spectroscopy,'' {\em Chemical Physics}, vol.~494, pp.~103--110, 2017.

\bibitem{SHOIFET2015}
E.~Shoifet, G.~Schulz, and C.~Schick, ``Temperature modulated differential
  scanning calorimetry – extension to high and low frequencies,'' {\em
  Thermochimica Acta}, vol.~603, pp.~227 -- 236, 2015.
\newblock Chip Calorimetry.

\bibitem{AgingGLY2019}
L.~A. Roed, T.~Hecksher, J.~C. Dyre, and K.~Niss, ``Generalized
  single-parameter aging tests and their application to glycerol,'' {\em The
  Journal of Chemical Physics}, vol.~150, no.~4, p.~044501, 2019.

\bibitem{harrison1976dynamic}
G.~Harrison, {\em The dynamic properties of supercooled liquids}.
\newblock Academic, London, 1976.

\bibitem{vdw2012GT}
B.~Jakobsen, T.~Hecksher, T.~Christensen, N.~B. Olsen, J.~C. Dyre, and K.~Niss,
  ``Communication: Identical temperature dependence of the time scales of
  several linear-response functions of two glass-forming liquids,'' {\em
  Journal of Chemical Physics}, vol.~136, no.~081102, 2012.

\bibitem{Stickel1996}
F.~Stickel, E.~W. Fischer, and R.~Richert, ``Dynamics of glass‐forming
  liquids. ii. detailed comparison of dielectric relaxation, dc‐conductivity,
  and viscosity data,'' {\em The Journal of Chemical Physics}, vol.~104, no.~5,
  pp.~2043--2055, 1996.

\bibitem{Hansen1997}
C.~Hansen, F.~Stickel, T.~Berger, R.~Richert, and E.~W. Fischer, ``Dynamics of
  glass-forming liquids. iii. comparing the dielectric alpha- and
  beta-relaxation of 1-propanol and o-terphenyl,'' {\em The Journal of Chemical
  Physics}, vol.~107, no.~4, pp.~1086--1093, 1997.

\bibitem{Kremer2018}
F.~Kremer and A.~Loidl, eds., {\em The Scaling of Relaxation Processes}.
\newblock Springer, 2018.

\bibitem{GT1994comp}
T.~Christensen and N.~B. Olsen, ``Comparative measurements of the electrical
  and shear mechanical response functions in some supercooled liquids,'' {\em
  Journal of Non-Crystalline Solids}, vol.~172-174, pp.~357--361, 1994.

\bibitem{Manda2001}
M.~Cutroni and A.~Mandanici, ``The alpha-relaxation process in simple glass
  forming liquid m-toluidine. ii. the temperature dependence of the mechanical
  response,'' {\em The Journal of Chemical Physics}, vol.~114, no.~16,
  pp.~7124--7129, 2001.

\bibitem{7liqs2005GT}
B.~Jakobsen, K.~Niss, and N.~B. Olsen, ``Dielectric and shear mechanical alpha
  and beta relaxations in seven glass-forming liquids,'' {\em Journal of
  Chemical Physics}, vol.~123, no.~234511, 2005.

\bibitem{GT1994HEX}
T.~Christensen and N.~B. Olsen, ``Quasistatic measurement of the
  frequenct-dependent bulk and shear modulus of supercooled liquids,'' {\em
  Journal of Non-Crystalline Solids}, vol.~172-174, pp.~362--364, 1994.

\bibitem{GT2013Mech}
T.~Hecksher, N.~B. Olsen, K.~A. Nelson, J.~C. Dyre, and T.~Christensen,
  ``Mechanical spectra of glass-forming liquids. i. low-frequency bulk and
  shear moduli of dc704 and 5-ppe measured by piezoceramic transducers,'' {\em
  The Journal of Chemical Physics}, vol.~138, no.~12, p.~12A543, 2013.

\bibitem{GT2014bulk}
D.~Gundermann, K.~Niss, T.~Christensen, J.~C. Dyre, and T.~Hecksher, ``The
  dynamic bulk modulus of three glass-forming liquids,'' {\em Journal of
  Chemical Physics}, vol.~140, no.~244508, 2014.

\bibitem{Roed2015PPE}
L.~A. Roed, K.~Niss, and B.~Jakobsen, ``Communication: High pressure specific
  heat spectroscopy reveals simple relaxation behavior of glass forming
  molecular liquid,'' {\em The Journal of Chemical Physics}, vol.~143, no.~22,
  p.~221101, 2015.

\bibitem{Casa2016}
R.~Casalini, S.~S. Bair, and C.~M. Roland, ``Density scaling and decoupling in
  o-terphenyl, salol, and dibutyphthalate,'' {\em The Journal of Chemical
  Physics}, vol.~145, no.~6, p.~064502, 2016.

\bibitem{ReinerZorn1997}
R.~Zorn, F.~I. Mopsik, G.~B. McKenna, L.~Willner, and D.~Richter, ``Dynamics of
  polybutadienes with different microstructures. 2. dielectric response and
  comparisons with rheological behavior,'' {\em The Journal of Chemical
  Physics}, vol.~107, no.~9, pp.~3645--3655, 1997.

\bibitem{GT2008monoalc}
B.~Jakobsen, C.~Maggi, T.~Christensen, and J.~C. Dyre, ``Investigation of the
  shear-mechanical and dielectric relaxation processes in two monoalcohols
  close to the glass transition.,'' {\em Journal of Chemical Physics},
  vol.~129, no.~184502, 2008.

\bibitem{C7CP06482A}
M.~H. Jensen, C.~Gainaru, C.~Alba-Simionesco, T.~Hecksher, and K.~Niss, ``Slow
  rheological mode in glycerol and glycerol–water mixtures,'' {\em Phys.
  Chem. Chem. Phys.}, vol.~20, pp.~1716--1723, 2018.

\bibitem{gabriel2020GLY}
J.~P. Gabriel, P.~Zourchang, F.~Pabst, A.~Helbling, P.~Weigl, T.~B{\"o}hmer,
  and T.~Blochowicz, ``Intermolecular cross-correlations in the dielectric
  response of glycerol,'' {\em Physical Chemistry Chemical Physics}, vol.~22,
  no.~20, pp.~11644--11651, 2020.

\bibitem{THKN2018}
K.~Niss and T.~Hecksher, ``Perspective: Searching for simplicity rather than
  universality in glass-forming liquids,'' {\em The Journal of Chemical
  Physics}, vol.~149, no.~23, p.~230901, 2018.

\bibitem{GT2008cryo}
B.~Igarashi, T.~Christensen, E.~H. Larsen, N.~B. Olsen, I.~H. Pedersen,
  T.~Rasmussen, and J.~C. Dyre, ``A cryostat and temperature control system
  optimized for measuring relaxations of glass-forming liquids,'' {\em Review
  of Scientific Instruments}, vol.~79, no.~045105, pp.~1--13, 2008.

\bibitem{GT2008imped}
B.~Igarashi, T.~Christensen, E.~H. Larsen, N.~B. Olsen, I.~H. Pedersen,
  T.~Rasmussen, and J.~C. Dyre, ``An impedance-measurement setup optimized for
  measuring relaxations of glass-forming liquids,'' {\em Review of Scientific
  Instruments}, vol.~79, no.~4, p.~045106, 2008.

\bibitem{GT1995shear}
T.~Christensen and N.~B. Olsen, ``A rheometer for the measurement of a high
  shear modulus covering more than seven decades of frequency below 50 khz,''
  {\em Review Scientific Instruments}, vol.~66, no.~5019, 1995.

\bibitem{GT1994bulk}
T.~Christensen and N.~B. Olsen, ``Determination of the frequency-dependent bulk
  modulus of glycerol using a piezoelectric spherical shell,'' {\em Physical
  Review B}, vol.~49, no.~15396, 1994.

\bibitem{GT2012CTE}
K.~Niss, D.~Gundermann, T.~Christensen, and J.~C. Dyre, ``Dynamic thermal
  expansivity of liquids near the glass transition,'' {\em Phys. Rev. E},
  vol.~85, p.~041501, Apr 2012.

\bibitem{GT2008speche}
T.~Christensen, N.~B. Olsen, and J.~C. Dyre, ``Can the frequency-dependent
  isobaric specific heat be measured by thermal effusion methods?,'' {\em AIP
  Conference Proceedings}, vol.~982, no.~138, 2008.
\newblock Paper presented at the Fifth International Workshop on Complex
  Systems in Sendai 2007.

\bibitem{GT2010speche}
B.~Jakobsen, N.~B. Olsen, and T.~Christensen, ``Frequency-dependent specific
  heat from thermal effusion in spherical geometry,'' {\em Physical Review E},
  vol.~81, no.~061505, 2010.

\bibitem{HEXdiel}
M.~Nakanishi and R.~Nozaki, ``Dynamics and structure of hydrogen-bonding glass
  formers: Comparison between hexanetriol and sugar alcohols based on
  dielectric relaxation,'' {\em Physical Review E}, vol.~81, p.~041501, 2010.

\bibitem{GaT2017FitABSQLN}
T.~Hecksher, N.~B. Olsen, and J.~C. Dyre, ``Model for the alpha and beta
  shear-mechanical properties of supercooled liquids and its comparison to
  squalane data,'' {\em The Journal of Chemical Physics}, vol.~146, no.~15,
  p.~154504, 2017.

\bibitem{FitParabol2010}
Y.~S. Elmatad, D.~Chandler, and J.~P. Garrahan, ``Corresponding states of
  structural glass formers. ii,'' {\em The Journal of Physical Chemistry B},
  vol.~114, no.~51, pp.~17113--17119, 2010.
\newblock PMID: 21138279.

\bibitem{Blochowicz2016}
T.~Blochowicz, C.~Gainaru, P.~Medick, C.~Tschirwitz, and E.~A. Rössler, ``The
  dynamic susceptibility in glass forming molecular liquids: The search for
  universal relaxation patterns ii,'' {\em The Journal of Chemical Physics},
  vol.~124, no.~13, p.~134503, 2006.

\bibitem{GaT2010FitABdiel}
N.~Sağlanmak, A.~I. Nielsen, N.~B. Olsen, J.~C. Dyre, and K.~Niss, ``An
  electrical circuit model of the alpha-beta merging seen in dielectric
  relaxation of ultraviscous liquids,'' {\em The Journal of Chemical Physics},
  vol.~132, no.~2, p.~024503, 2010.

\end{thebibliography}

	\appendix
	\beginsupplement

%

\section{Description of experimental methods}
A piezoelectric shear modulus gauge \cite{GT1995shear} was used for measurements of the dynamic shear-mechanical reponse of hexanetriol and squalane. The device consists of  three piezoceramic disks that are centered towards another with a gap inbetween that can be filled with a liquid material. Each disk is a capacitor due to electrodes on the surface of each side of the disks, that will show a lower capacitance when mechanically clamped by the inserted liquid, as the capacity of the priezceramic material is dependent on its strain state. The three disks are electrically connected in a manner that results in a counter-acting change of surface area, e.g. an expansion of the outer pair of disks is accompanied by the contraction of the inner disk. Applying an alternating voltage thus leads to a strain field within the liquid that oscillates with time. Based on a model assuming pure shear deformation of the liquid at the edges of the disks, the shear dynamic shear modulus $G(\omega)$ is deduced from measurements in the quasi-static mode. The frequencies covered range form \SI{1}{\milli\hertz} to \SI{50}{\kilo\hertz}, as resonances occur at higher frequencies. The dynamic range lies between \SI{1e5}{\pascal} and \SI{1e10}{\pascal}.\\

For measurements of the dynamic bulk-mechanical modulus of the two supercooled liquids, a piezoelectric bulk modulus gauge was used \cite{GT1994bulk,GT1994HEX,GT2014bulk}. The device consists of a sperical shell of piezoceramic material that contracts and expands with an applied alternating voltage. Electrodes on the inner and outer surfaces of the shell allow for measuring its frequency-dependent capacitance. As the capacitance is connected to the strain state of the piezoceramic material of the shell as identified by means of an electric equivalent circuit model, the difference in the response of an empty shell and a liquid-filled shell gives access to the bulk-mechanical properties of the liquid. The frequency range of the device is within \SI{1}{\milli\hertz} and \SI{10}{\kilo\hertz}.\\

The frequency-dependent dielectric permittivity was measured using a parallel-plate capacitor with diameter of \SI{10}{\milli\metre} and plate distance of \SI{50}{\micro\metre} ensured by four \SI{1}{\milli\meter\squared} Polyimide spacers assumed as being rigid. Capacitance recorded using the electrical equipment described in Ref.~\onlinecite{GT2008imped} within  frequency range of \SI{1}{\milli\hertz} to \SI{1}{\mega\hertz}.\\

For measurements of the specific heat, a cup filled with liquid with a bead in the middle of the volume was used. The bead consists of a negative temperature-coefficient thermistor that serves both as a a temperature-gauge and a heating device \cite{GT2008speche,GT2010speche}. By introducing a specific amount of thermal energy into the system in terms of a dynamic  power $P(\omega)$ and measuring the nonlinear periodic temperature.change at the surface of the bead, the effusivity of the material can be determined. It characterizes the material's ability to exchange thermal energy with its surroundings, i. e. the thermal effusivity, that is connected to the dynamic specific heat via the thermal conductance. The method is based on the evaluation of higher order contributions to the measured thermal impedance using an iterative solution technique, which emphasizes the need for highly sensitive measurements. By fitting the higher-order thermal impedance to a suitable electrical-equivalent circuit model, the thermal impedance of the liquid can be derived and the spectra of the specific heat of the liquid be determined.\\

Measurements of the thermal expansion were performed by means of a parallel-plate capacitor in which the sample material dictates the spacing of the two capacitor's plates. Upon a change in temperature the volume of the sample material de- or increases, which is reflected in the measured capacitance. Based on highly-accurate linear temperature jumps the timescale on which thermal expansion occurs is measured as described previously \cite{GT2012CTE}.

\begin{figure}[h!]
	\centering
	\includegraphics[width=\columnwidth]{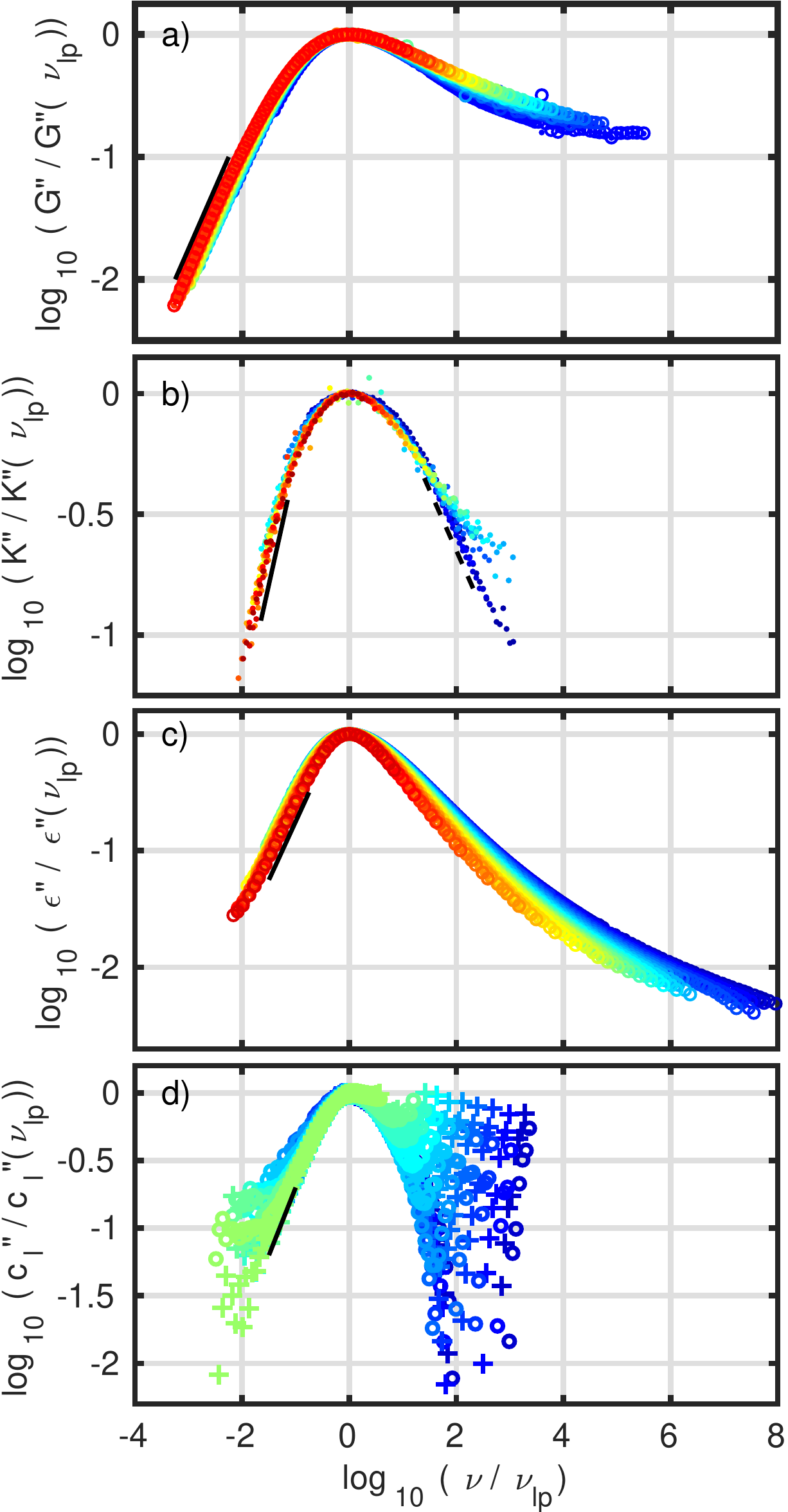}
	\caption{Normalized loss contribution of the dynamic complex response functions measured on the equilibrium supercooled liquid state of 1,2,6-Hexanetriol as functions of normalized frequency. The normalization is based on the loss peak position $\nu_{lp}$. Open symbols depict experimental data: \textbf{a)} shear modulus $G'' (\omega)$, \textbf{b)} bulk modulus $K'' (\omega)$, \textbf{c)} dielectric permittivity $\varepsilon'' (\omega)$, and \textbf{d)} specific heat $c_l'' (\omega)$. Solid black lines depict power-law behavior with slope \num{1} and dashed black lines powerlaw behavior with slope \num{-0.5}.}
	\label{FigTTS}
\end{figure}

\section{Normalized spectral data}
In this seciton, loss data of the complex response functions measured on hexanetriol and squalane are presented after normalization to the individual peak position to facilitate the comparison of temperature-specific spectral shapes, i.e. time-temperature superposition.

\subsection{Hexanetriol}
Shear-mechanical spectra are plotted normalized to their peak-position in fig.~\ref{FigTTS}a).  The data show a convincing collapse on the low-frequency flank, were a power-law behavior with a slope of \num{1} is observed. On the high-frequency flank, the powerlaw deviates from that expected for a pure primary relaxation signal, indicating a slight impact of the $\beta$-process in form of a shoulder in the spectra.

Bulk-mechanical spectra were normalized to the position of the primary relaxation contribution determined from the overall loss peak maximum and plotted in Fig.~\ref{FigTTS}b). The spectra coincide for all measured tempratures with exception of the lowest temperatures, for which a broadening on the low-frequency flank is observed. Spectra measured at higher temperatures show power-law behavior on the low-frequency flank with an exponent of \num{1} (solid line in Fig.~\ref{FigTTS}b)). Only on the high-frequency flank of spectra measured at mediocre temperatures values above the expected power law of \num{-1/2} are observed (dashed line in Fig.~\ref{FigTTS}b)).

In Fig.~\ref{FigTTS}c) the dielectric spectra are plotted normalized to the loss peak position based on the timescales determined from the overall loss peak maximum.Spectra do not show a distinct secondary process in the measured frequency-regime.

The loss contribution of heat capacity data normalized to the peak position is plotted in Fig.~\ref{FigTTS}d). As discussed in the manuscript, the spectra show strong scattering on the low- and high-frequency flank that prohibit a detailed discussion of the peak shape. However, for the measured temperatures the data collapses in a wider range around the peak.

\subsection{Squalane}
\begin{figure*}[h!]	
	\begin{minipage}{\columnwidth}
		\includegraphics[width=\textwidth]{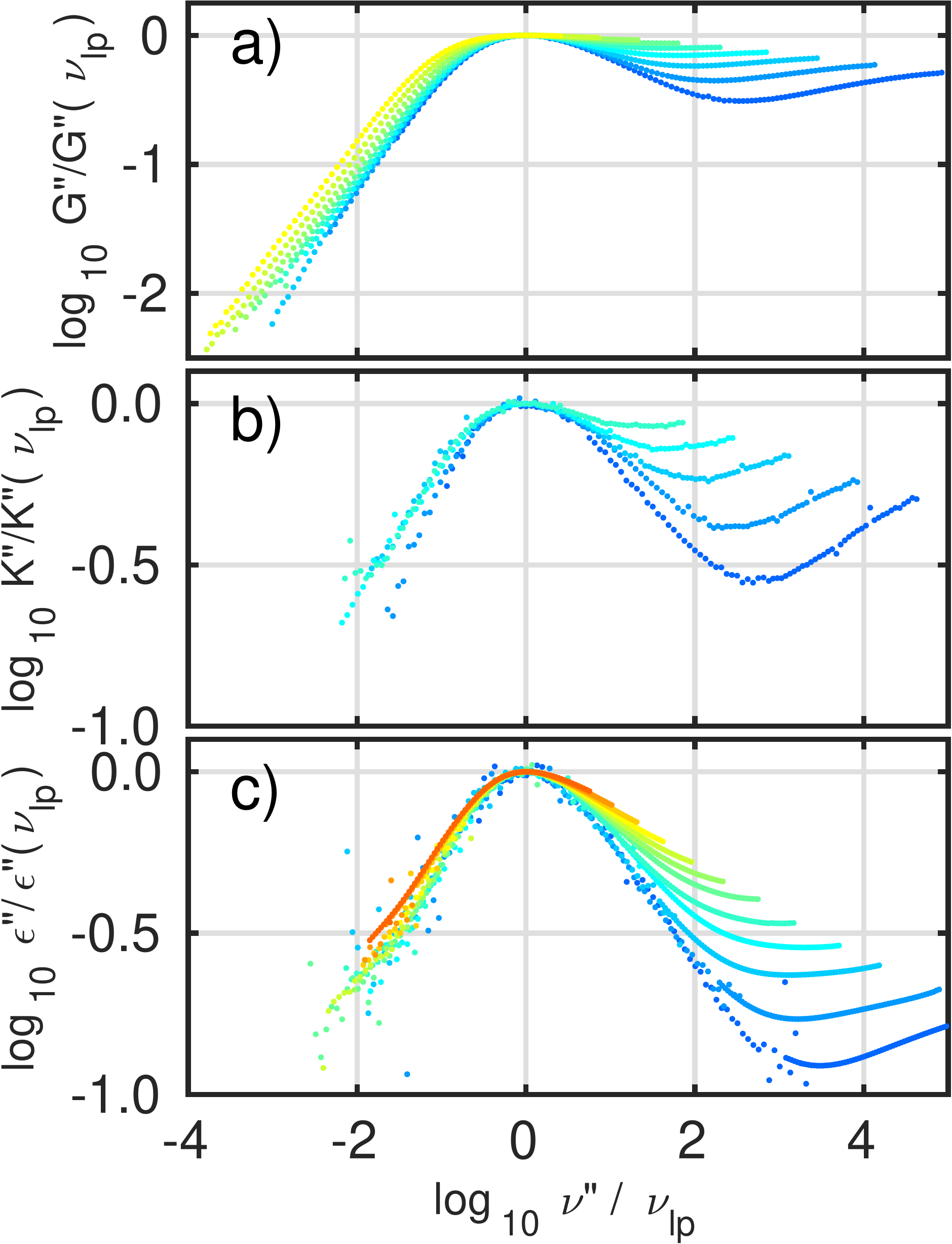}
	\end{minipage}	
	\begin{minipage}{0.83\columnwidth}
		\includegraphics[width=\textwidth]{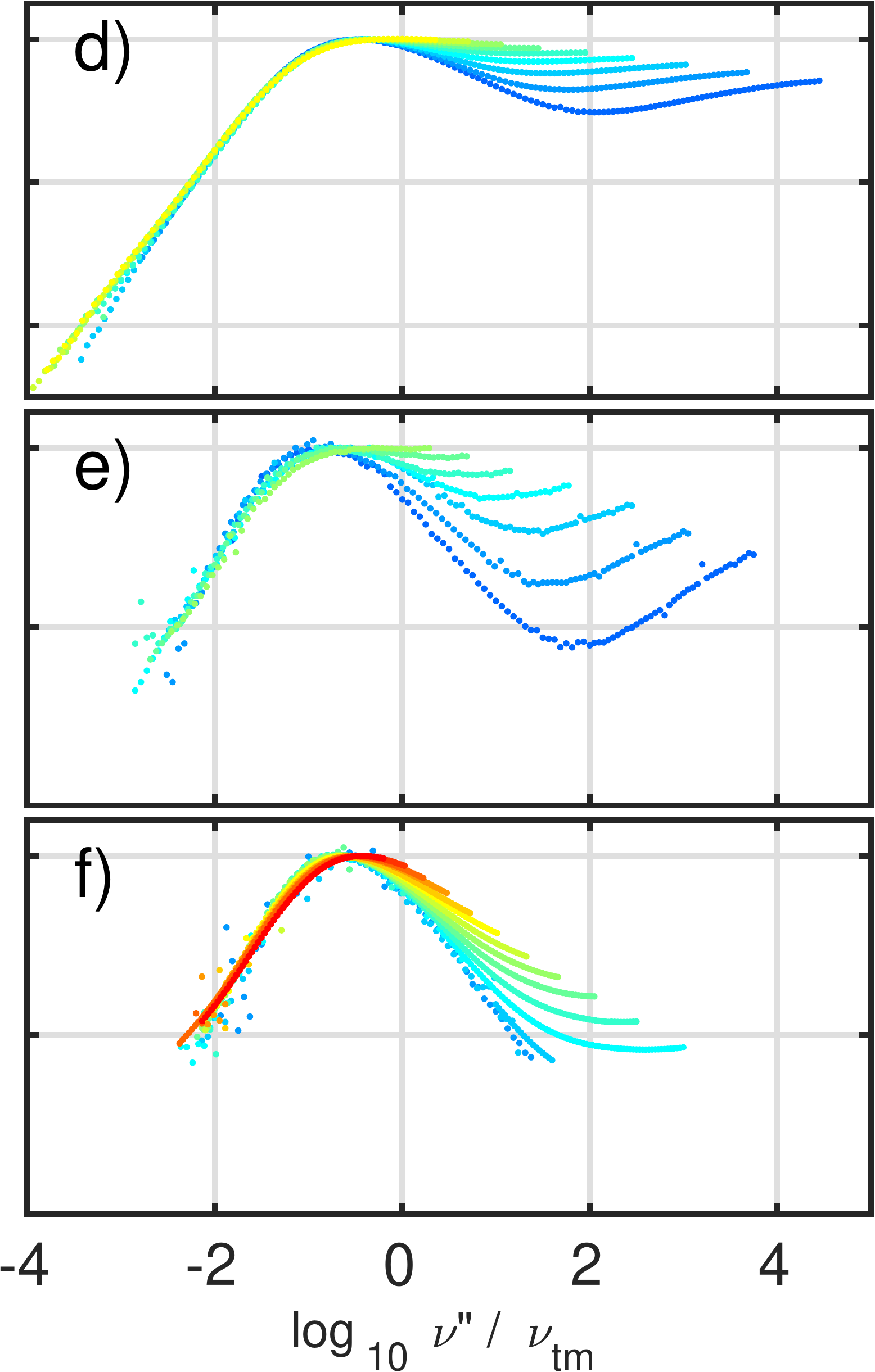}
	\end{minipage}
	
	\caption{Normalized loss contribution of the dynamic shear modulus $G'' (\omega)$ of the supercooled liquid state of Squalane as a function of normalized frequency. The normalization is based on the timescale determined from the overall loss peak maximum (\textbf{a)} to \textbf{c)}) and the timescale based on the terminal mode (\textbf{d)} to \textbf{f)}) with indicators for powerlaw-behavior with exponents of \num{1} (solid line) and \num{-0.5} (dashed black line).}
	
	\label{FigSQLNtts}
\end{figure*}

In Fig.~\ref{FigSQLNtts}, panel a) to c), the loss contributions of shear-mechanical, bulk-mechanical, and dielectric data are plotted after normalization to the peak position. As spectra measured at elevated temperature show a stong merging of the primary and secondary relaxation contribution, the time scales connected to the overall loss-peak position can not be assumed as reflecting solely the process of structural relaxation.

Therefore, in Fig.~\ref{FigSQLNtts} panel d) to f), the spectra are normalized to the terminal mode by a manually achieved collapse of the data on the low-frequency flank. Time scales are derived from the applied shift relative to the peak-position-related time scale at the lowest temperature.
\newpage

\onecolumngrid
\clearpage
\section{Electrical-equivalent circuit models}
In this section, the models connected to shear-mechanical, bulk-mechanical, and dielectric response are presented with a derivation of the model-based expressions.
\subsection{Model for dynamic shear-mechanical response}
A model for the primary and secondary relaxation in shear-mechanical data was compared to Squalane-data by Hecksher et al. \cite{GaT2017FitABSQLN}. It is based on an electrical-equivalent-circuit that holds elements reflecting the dynamic shear-mechanical properties  of  a  supercooled  liquid assuming additivity of $\alpha$- and $\beta$-contributions. The  alpha  process  is  represented  by  a  standard  RC-element corresponding  to  the  Maxwell  model in  parallel  with  a  Cole-Cole  retardation  element with exponent 1/2. The beta process is represented by an additional CCRE.
\begin{figure}[h!]
	\includegraphics[width=0.3\textwidth]{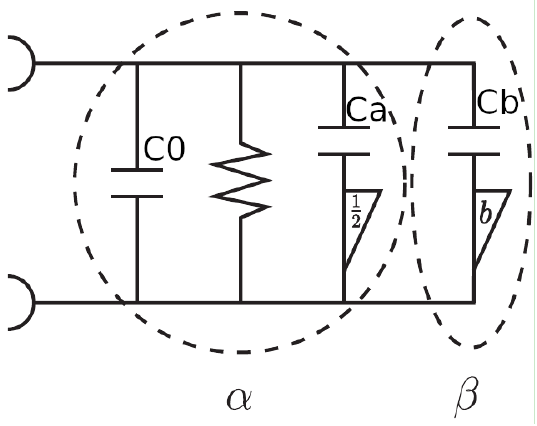}
	\caption{Electrical-equivalent-circuit model for the dynamic shear-mechanical properties  of  a  supercooled  liquid from Ref.~\onlinecite{GaT2017FitABSQLN}.}\label{FitABshearCircuit}
\end{figure}\\
The model leads to the following expression:
\begin{equation}\label{ABmodelShear}
	G(\omega) = (J(\omega))^{-1} = \left( J_a \left( 1 + \frac{1}{i \omega \tau_a} + \frac{k_{a} / J_a}{1 + k_{a}/c_a (i \omega \tau_a)^{1/2}} \right) + \frac{J_b}{1 + (i \omega \tau_b)^{b}} \right)^{-1}
\end{equation}\\
The expression holds seven parameters:   $J_a$ in \si{\per\giga\per\pascal}, $\tau_a$ in \si{\second}, $k_{a}$, $c_{a}$, $J_b$ in  \si{\per\giga\per\pascal}, $\tau_b$ in \si{\second}, and $b$.\\
\\
Starting from the individual components of the electrical-equivalent circuit model as shown in fig.~\ref{FitABshearCircuit}:
\begin{equation}\label{ABmodelShearApp1}
	C(\omega) = C_0 + \frac{1}{i \omega R} + \left(\frac{1}{C_a} + K_a (i \omega)^{a} \right)^{-1} + \left(\frac{1}{C_b} + K_b (i \omega)^{b} \right)^{-1}\\
\end{equation}
With $\tau_x = (q_x K_x)^{1/x}$, and specifically $\tau_a = R \cdot C_x$:
\begin{equation}\label{ABmodelShearApp2}
	\begin{aligned}
		C(\omega) &= C_0 + \frac{C_x}{i \omega \tau_a} + \left(\frac{1}{C_a} + \frac{1}{q_a} (i \omega \tau_a)^{a} \right)^{-1} + \left(\frac{1}{C_b} + \frac{1}{q_b} (i \omega \tau_b)^{b} \right)^{-1}\\
		&= C_0 \left( 1 + \frac{C_x / C_0}{i \omega \tau_a} + \frac{C_a / C_0}{1 + \frac{C_a}{q_a} (i \omega \tau_a)^{a}} \right) + \frac{C_b}{1 + \frac{C_b}{q_b} (i \omega \tau_b)^{b}}\\
	\end{aligned}
\end{equation}\\
With $J_a = C_0 / c$, $J_b = C_b / c$, $j_a = C_a / c$:
\begin{equation}\label{ABmodelShearApp3}
	J(\omega)= J_a \left( 1 + \frac{j_x / J_a}{i \omega \tau_a} + \frac{j_a / J_a}{1 + j_a \cdot c / q_a (i \omega \tau_a)^{a}} \right) + \frac{J_b}{1 + J_b \cdot c / q_b (i \omega \tau_b)^{b}}
\end{equation}\\
With $k_{a0} = j_x / J_a$, $k_{a} = j_a$, $c_a = q_a/ c$, $k_{b} = J_b  \cdot c / q_b$:
\begin{equation}\label{ABmodelShearApp4}
	J(\omega)= J_a \left( 1 + \frac{k_{a0}}{i \omega \tau_a} + \frac{k_{a} / J_a}{1 + k_{a}/c_a (i \omega \tau_a)^a} \right) + \frac{J_b}{1 + k_{b} (i \omega \tau_b)^{b}}
\end{equation}\\
Finally, with $C_x = C_0$ resulting in $k_{a0} = 1$, as well as $a = 1/2$ and $k_{b} = 1$:
\begin{equation}\label{ABmodelShearAppFin5}
	J(\omega)= J_a \left( 1 + \frac{1}{i \omega \tau_a} + \frac{k_{a} / J_a}{1 + k_{a}/c_a (i \omega \tau_a)^{1/2}} \right) + \frac{J_b}{1 + (i \omega \tau_b)^{b}}
\end{equation}

\subsection{Model for dynamic bulk-mechanical response}
\begin{figure}[h!]
	\includegraphics[width=0.35\textwidth]{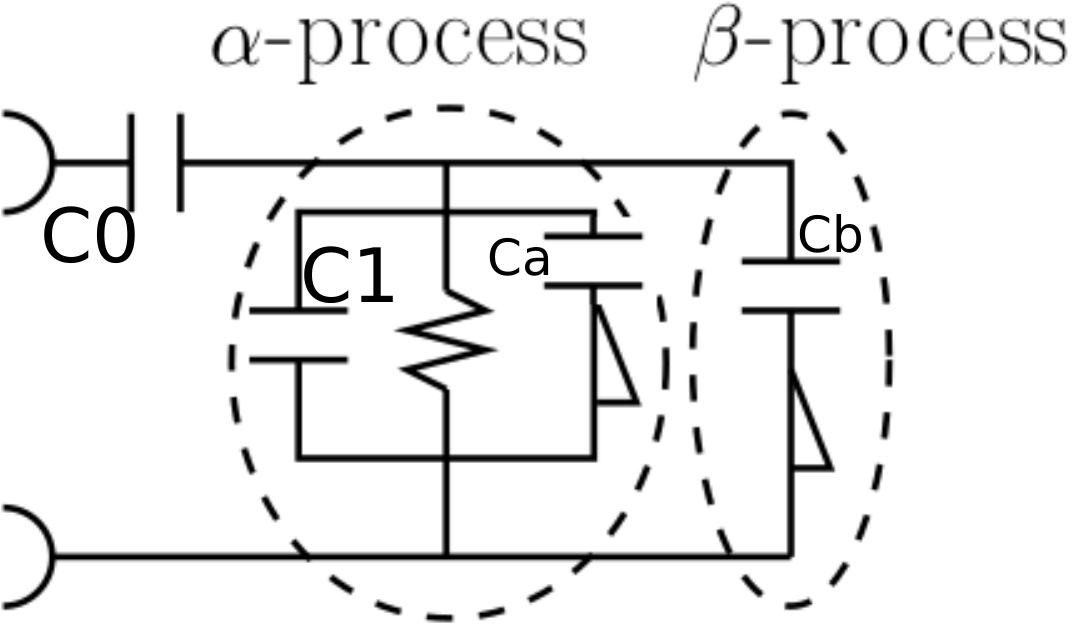}
	\caption{Electrical-equivalent-circuit model for the bulk-mechanical dielectric properties  of  a  supercooled  liquid.}\label{FitABbulkCircuit}
\end{figure}

In order to create a model that describes the bulk-mechanical properties of a supercooled liquid, the shear-related electrical-equivalent-circuit model was extended by an additional capacitor as proposed in \cite{GaT2010FitABdiel} and shown in fig.~\ref{FitABbulkCircuit}. The capacitor characterizes the bulk-mechanical response in the limit of zero frequency.\\
\\
This model leads to the following expression:
\begin{equation}\label{ABmodelBulk}
	K(\omega)= K_0 + \left[ J_a \left( 1 + \frac{1}{i \omega \tau_a} + \frac{k_{a}/J_{a}}{1 + k_{a}/c_{a} (i \omega \tau_a)^{1/2}} \right) + \frac{J_b}{1 + (i \omega \tau_b)^{b}} \right]^{-1}
\end{equation}
The expression holds eight parameters: $K_0$ in \si{\giga\pascal}, $J_a$ in \si{\per\giga\per\pascal}, $\tau_a$ in \si{\second}, $k_{a}$, $c_{a}$, $J_b$ in  \si{\per\giga\per\pascal}, $\tau_b$ in \si{\second}, and $b$.\\
\\
Starting from the individual components of the electrical-equivalent circuit model as shown in fig.~\ref{FitABbulkCircuit}:

\begin{equation}\label{ABmodelBulkApp1}
	C(\omega) = \left( \frac{1}{C_0} + \left[ C_1 +  \frac{1}{i \omega R} + \left(\frac{1}{C_a} + K_a (i \omega)^{a} \right)^{-1} + \left(\frac{1}{C_b} + K_b (i \omega)^{b} \right)^{-1}  \right]^{-1} \right)^{-1}\\
\end{equation}
With $\tau_x = (q_x K_x)^{1/x}$, and specifically $\tau_a = R \cdot C_x$:
\begin{equation}\label{ABmodelBulkApp2}
	C(\omega) = \left( \frac{1}{C_0} + \left[ C_1 +  \frac{C_x}{i \omega \tau_a} + \frac{C_a}{1 + \frac{C_a}{q_a} (i \omega \tau_a)^{a}} + \frac{C_b}{1 + \frac{C_b}{q_b} (i \omega \tau_b)^{b}}  \right]^{-1} \right)^{-1}\\
\end{equation}
With $K(\omega) = 1 / C(\omega)$, $K_0 = c / C_0$, $J_a = C_1 / c$, $J_b = C_b / c$, $j_x = C_x / c$:
\begin{equation}\label{ABmodelBulkApp3}
	K(\omega)= K_0 + \left[ J_a \left( 1 + \frac{j_x / J_a}{i \omega \tau_a} + \frac{j_a / J_a}{1 + j_a \cdot c / q_a (i \omega \tau_a)^{a}} \right) + \frac{J_b}{1 + J_b \cdot c / q_b (i \omega \tau_b)^{b}} \right]^{-1}
\end{equation}\\
With $k_{a0} = j_x / J_a$, $k_{a} = j_a$, $c_{a} = q_a / c$, $k_{b} = J_b  \cdot c / q_b$:
\begin{equation}\label{ABmodelBulkApp4}
	K(\omega)= K_0 + \left[ J_a \left( 1 + \frac{k_{a0}}{i \omega \tau_a} + \frac{k_{a}/J_{a}}{1 + j_{a}/c_{a} (i \omega \tau_a)^a} \right) + \frac{J_b}{1 + k_{b} (i \omega \tau_b)^{b}} \right]^{-1}
\end{equation}\\
Finally, with $C_x = C_0$ resulting in $k_{a0} = 1$, as well as $a = 1/2$ and $k_{b} = 1$:
\begin{equation}\label{ABmodelBulkAppFin}
	K(\omega)= K_0 + \left[ J_a \left( 1 + \frac{1}{i \omega \tau_a} + \frac{k_{a}/J_{a}}{1 + k_{a}/c_{a} (i \omega \tau_a)^{1/2}} \right) + \frac{J_b}{1 + (i \omega \tau_b)^{b}} \right]^{-1}
\end{equation}

\subsection{Model for dynamic dielectric response}
\begin{figure}[h!]
	\includegraphics[width=0.5\textwidth]{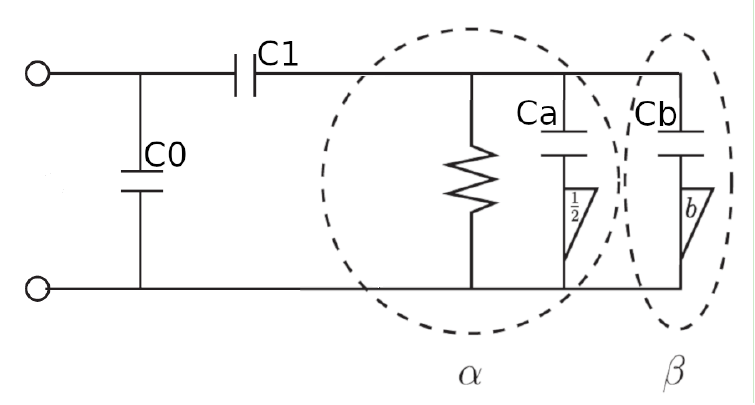}
	\caption{Alternative electrical-equivalent-circuit model for the dynamic dielectric properties of  a  supercooled  liquid.}\label{FitABdielCircuit3}
\end{figure}

To model the dielectric properties of a supercooled liquid, an electrical-equivalent-circuit model as depicted in fig.~\ref{FitABdielCircuit3} is used.\\
This model leads to the following expression:
\begin{equation}\label{ABmodelDiel3}
	\varepsilon(\omega) = \varepsilon_{\infty} + \frac{\Delta \varepsilon}{1 + \left( \frac{1}{i \omega \tau_a} + \frac{\varepsilon_a / \Delta \varepsilon}{1 + \varepsilon_a/k_a (i \omega \tau_a)^{1/2}} + \frac{\varepsilon_b / \Delta \varepsilon}{1 + (i \omega \tau_b)^{b}} \right)^{-1} }\\
\end{equation}\\
The expression holds eight parameters for the model: $\varepsilon_{\infty}$, $\Delta \varepsilon$, $\varepsilon_a$, $\tau_a$ in \si{\second}, $k_a$, $\varepsilon_b$, $\tau_b$ in \si{\second}, and $b$.\\
\\
Starting from the individual components of the electrical-equivalent circuit model as shown in fig.~\ref{FitABdielCircuit3}:
\begin{equation}\label{ABmodelDielApp1}
	C(\omega) = C_0 + \left( \frac{1}{C_1} + \left( \frac{1}{i \omega R} + \left(\frac{1}{C_a} + K_a (i \omega)^{a} \right)^{-1} + \left(\frac{1}{C_b} + K_b (i \omega)^{b} \right)^{-1} \right)^{-1} \right)^{-1}\\
\end{equation}
With $\tau_x = (q_x K_x)^{1/x}$, and specifically $\tau_a = R \cdot C_x$:
\begin{equation}\label{ABmodelDielApp2}
	\begin{aligned}
		C(\omega) &= C_0 + \left( \frac{1}{C_1} + \left( \frac{C_x}{i \omega \tau_a} + \left(\frac{1}{C_a} + \frac{1}{q_a} (i \omega \tau_a)^{a} \right)^{-1} + \left(\frac{1}{C_b} + \frac{1}{q_b} (i \omega \tau_b)^{b} \right)^{-1}  \right)^{-1} \right)^{-1}\\
		&= C_0 + \frac{C_1}{1 + \frac{C_1}{\frac{C_x}{i \omega \tau_a} + \frac{C_a}{1 + \frac{C_a}{q_a} (i \omega \tau_a)^{a}} + \frac{C_b}{1 + \frac{C_b}{q_b} (i \omega \tau_b)^{b}}}}\\
		&= C_0 + \frac{C_1}{1 + \left( \frac{C_x / C_1}{i \omega \tau_a} + \frac{C_a / C_1}{1 + \frac{C_a}{q_a} (i \omega \tau_a)^{a}} + \frac{C_b / C_1}{1 + \frac{C_b}{q_b} (i \omega \tau_b)^{b}} \right)^{-1} }\\
	\end{aligned}
\end{equation}\\
With $\varepsilon_{\infty} = C_0 / C_{geo}$, $\Delta \varepsilon = C_1 / C_{geo}$, $\varepsilon_x = C_x / C_{geo}$:
\begin{equation}\label{ABmodelDielApp3}
	\varepsilon(\omega)= \varepsilon_{\infty} + \frac{\Delta \varepsilon}{1 + \left( \frac{\varepsilon_x / \Delta \varepsilon}{i \omega \tau_a} + \frac{\varepsilon_a / \Delta \varepsilon}{1 + \frac{\varepsilon_a C_{geo}}{q_a} (i \omega \tau_a)^{a}} + \frac{\varepsilon_b / \Delta \varepsilon}{1 + \frac{\varepsilon_b C_{geo}}{q_b} (i \omega \tau_b)^{b}} \right)^{-1} }\\
\end{equation}\\

With $k_{a,0} = \varepsilon_x / \Delta \varepsilon$, $ k_a = q_a/C_{geo}$, $k_b = \varepsilon_b C_{geo}/q_b$:
\begin{equation}\label{ABmodelDielApp4}
	\varepsilon(\omega) = \varepsilon_{\infty} + \frac{\Delta \varepsilon}{1 + \left( \frac{k_{a,0}}{i \omega \tau_a} + \frac{\varepsilon_a / \Delta \varepsilon}{1 + \varepsilon_a/k_a (i \omega \tau_a)^{a}} + \frac{\varepsilon_b / \Delta \varepsilon}{1 + k_b (i \omega \tau_b)^{b}} \right)^{-1} }\\
\end{equation}\\
Finally, with $C_x = C_0$ resulting in $l_{a0} = 1$, as well as $a = 1/2$ and $l_{b} = 1$:
\begin{equation}\label{ABmodelDielAppFin}
	\varepsilon(\omega) = \varepsilon_{\infty} + \frac{\Delta \varepsilon}{1 + \left( \frac{1}{i \omega \tau_a} + \frac{\varepsilon_a / \Delta \varepsilon}{1 + \varepsilon_a/k_a (i \omega \tau_a)^{1/2}} + \frac{\varepsilon_b / \Delta \varepsilon}{1 + (i \omega \tau_b)^{b}} \right)^{-1} }\\
\end{equation}

\clearpage
\twocolumngrid
\section{Spectral data  with EEC-model fits}
Spectral data sets for mechanical and dielectric data measured on hexanetriol and squalane are plotted in this section together with their respective by EEC-models and details on fitting parameters.

\subsection{1,2,6-Hexanetriol}
\begin{figure}[h!]
	\includegraphics[width=\columnwidth]{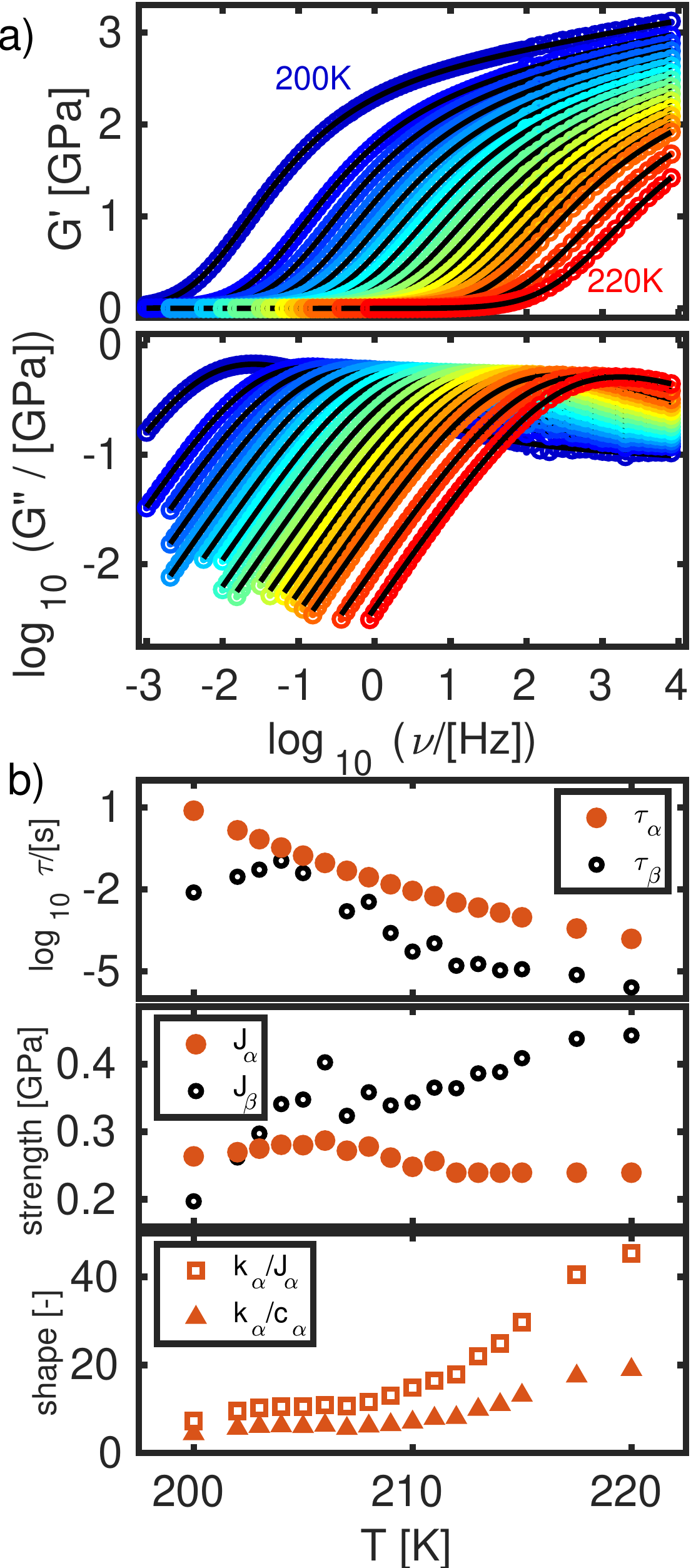}
	\caption{\textbf{a)} Storage and loss spectra of the dynamic shear modulus $G (\omega) = G' (\omega) + i G'' (\omega)$ of the supercooled liquid state of hexanetriol at selected temperatures ranging from \SI{200}{\kelvin} to \SI{220}{\kelvin}. After quenching to \SI{200}{\kelvin}, the shear modulus was measured in steps of \SI{1}{\kelvin} from \SI{202}{\kelvin} to \SI{214}{\kelvin} and steps of \SI{2.5}{\kelvin} from \SI{215}{\kelvin} to \SI{220}{\kelvin}. Points correspond to the first and open circles to the second, subsequently measured spectrum. Black lines correspond to fits of the data to the shear-mechanical EEC-model. \textbf{b)} Fit parameters as described in eq.~\ref{ABmodelShear}.}
	\label{FigHEXshearParams}
\end{figure}

\textbf{Shear-mechanical data} were fitted to the model applicable to dynamic shear-mechanical data as described by eq.~\ref{ABmodelShear}. It holds seven parameters to describe the response assuming a primary and secondary relaxation process. For the fitting, only one parameter was fixed to a constant (exponent of the $\beta$-relaxation, $b =$ 0.15), while the other six parameters were free. An exception was made for fits with the data of the highest four temperatures, which lack data at the high-frequency wing for a meaningful fit with free parameters. Here, the parameter $J_a$ was fixed to the $J_a$-value at the highest temperature for which a meaningful fit could still be gained from free parameters (see Fig.~\ref{FigHEXshearParams}).\\

\begin{figure}[h!]
	\includegraphics[width=\columnwidth]{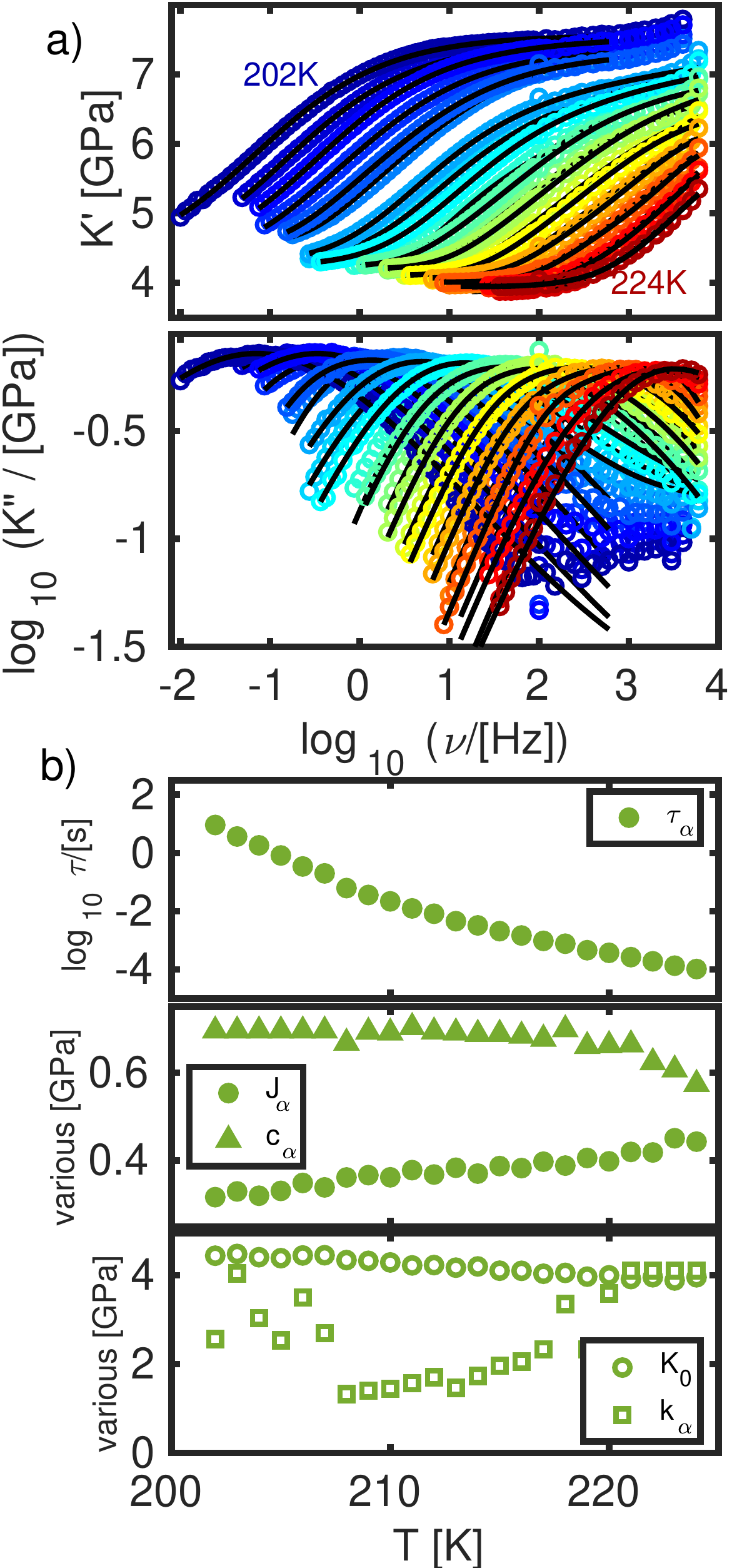}
	\caption{\textbf{a)} Storage and loss spectra of the dynamic bulk modulus $K (\omega) = K' (\omega) + i K'' (\omega)$ of the supercooled liquid state of hexanetriol for selected temperatures between \SI{202}{\kelvin} and \SI{224}{\kelvin}. Spectra were measured from \SI{223}{\kelvin} to \SI{203}{\kelvin} in steps of \SI{-2}{\kelvin} and from \SI{202}{\kelvin} to \SI{224}{\kelvin} in steps of \SI{+2}{\kelvin}. Black lines correspond to fits of the data to the shear-mechanical EEC-model. \textbf{b)} Fit parameters as described in eq.~\ref{ABmodelBulk}. Fits do not include secondary process.}
	\label{FigHEXbulkParams}
\end{figure}

\textbf{Bulk data} were fitted to eq.~\ref{ABmodelBulk} with $J_b$ set to zero, so that the secondary process was excluded from fitting in contrast to the shear-mechanical data. The parameter $c_a$ had to be restricted to a fixed value for the lowest six temperatures, $k_a$ was restricted to a fixed value for the highest three temperatures. Data measured at the other 14 temperatures were fitted with eight free parameters (see Fig.~\ref{FigHEXbulkParams}).\\

\textbf{Dielectric data} were fitted to the expression of the relevant EEC-model (eq.~\ref{ABmodelDiel3}), where the secondary relaxation contribution was excluded as in the case of the bulk-mechanical data. All remaining five parameters were kept free at all temperatures to obtain the fits.
\begin{figure}[h!]
	\includegraphics[width=\columnwidth]{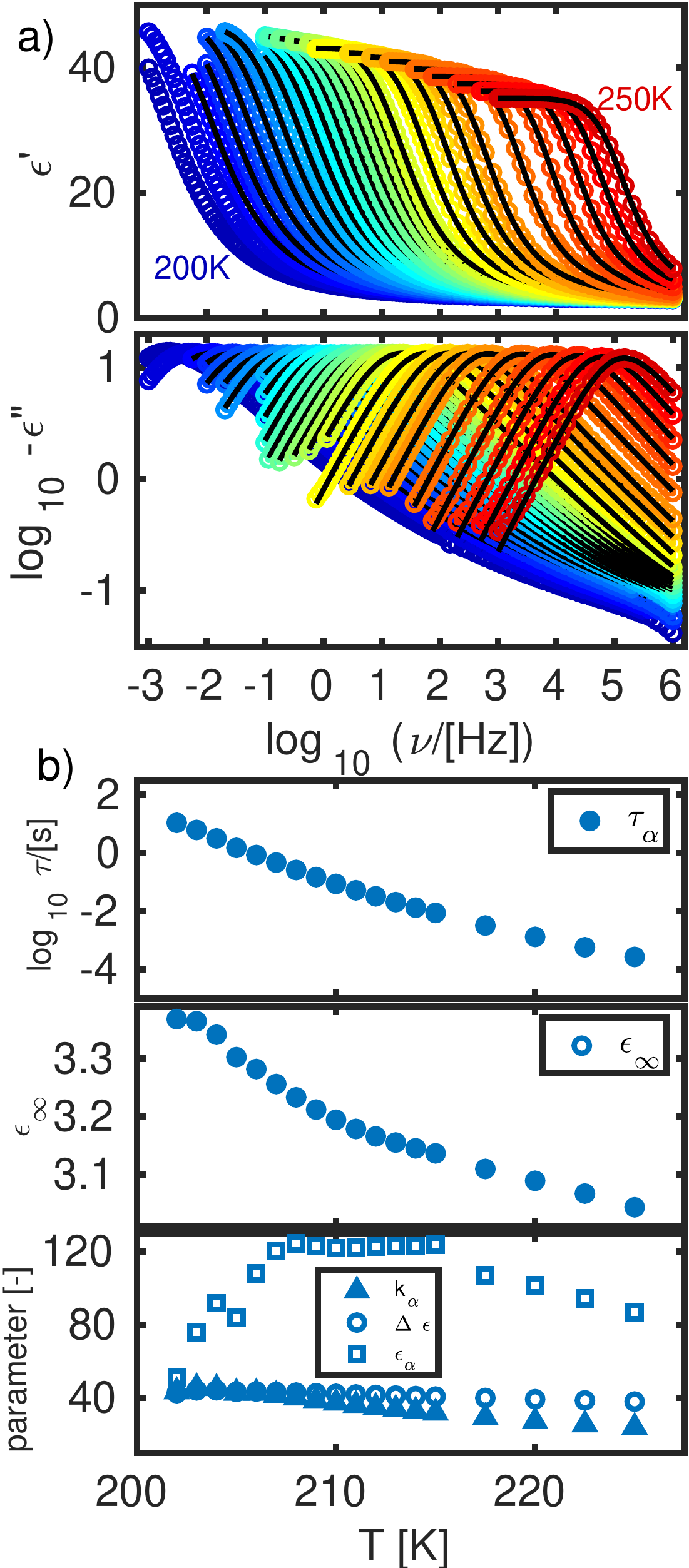}
	\caption{\textbf{a)} Storage and loss spectra of the dynamic dielectric permittivity $\varepsilon (\omega) = \varepsilon' (\omega) - i \varepsilon'' (\omega)$ of the supercooled liquid state of hexanetriol at selected temperatures ranging from \SI{200}{\kelvin} to \SI{250}{\kelvin}. After a quench to \SI{200}{\kelvin}, spectra were measured in temperature steps of \SI{1}{\kelvin} up to \SI{215}{\kelvin}, steps of \SI{2.5}{\kelvin} from \SI{217.5}{\kelvin} to \SI{225}{\kelvin}, and steps of \SI{5}{\kelvin} from \SI{230}{\kelvin} to \SI{250}{\kelvin}. Points correspond to the first, open circles to the second, subsequently measured spectrum. \textbf{b)} Fit parameters as described in eq.~\ref{ABmodelDiel3}. Fits do not include secondary process.}
	\label{FigHEXdielParams}
\end{figure}

\clearpage
\subsection{Squalane}
\begin{figure}[h!]
	\includegraphics[width=\columnwidth]{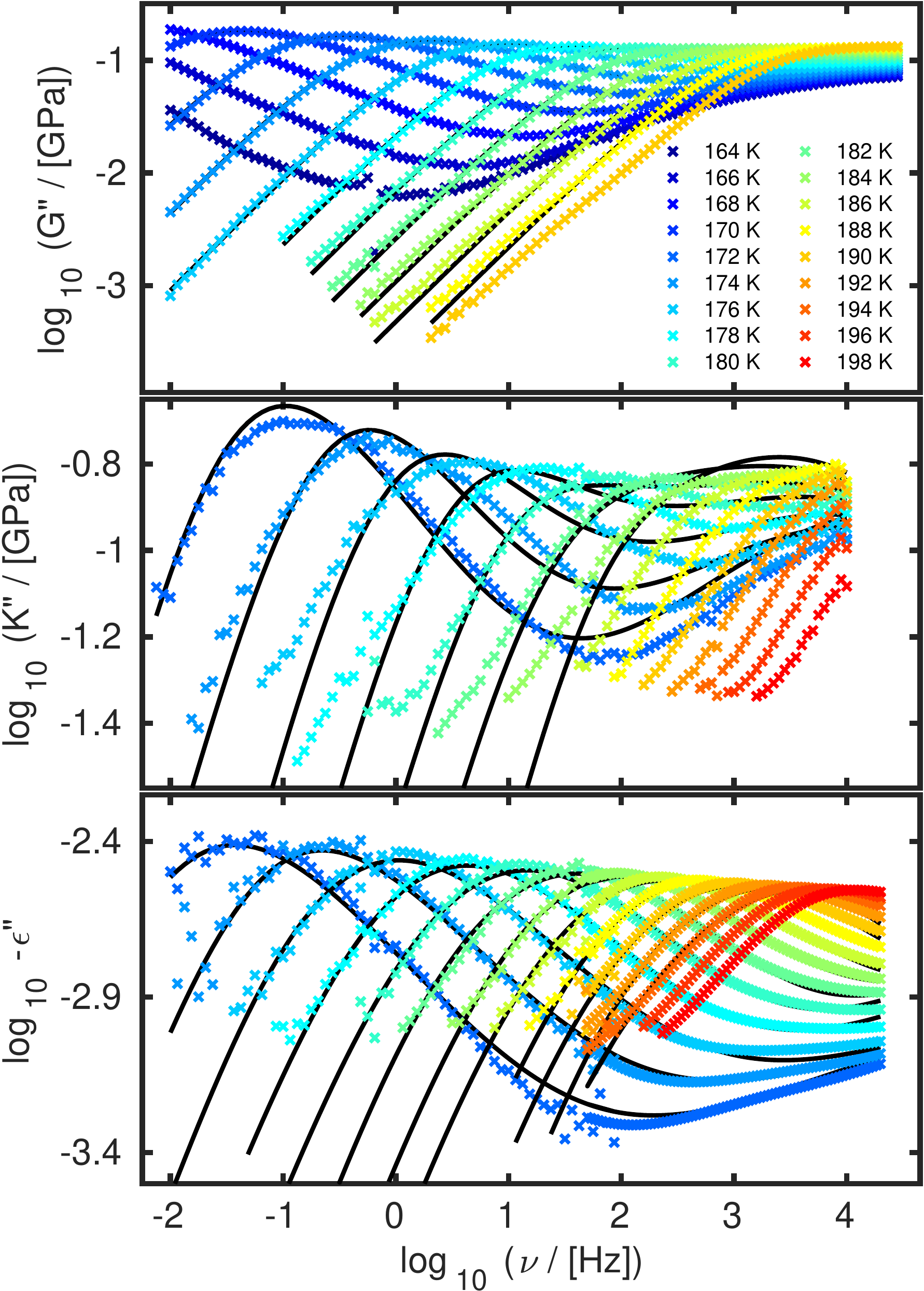}
	\caption{Loss spectra of the dynamic shear modulus $G (\omega)$, bulk modulus $K (\omega)$, and dielectric permittivity $\varepsilon(\omega)$ of squalane from \SI{164}{\kelvin} to \SI{190}{\kelvin} for shear-mechanical data, and from \SI{172}{\kelvin} to \SI{198}{\kelvin} for dielectric and bulk-mechanical data.}
	\label{App_FigSQLN}
\end{figure}

\newpage

\begin{figure}[h!]
	\includegraphics[trim = 0cm 0cm 0cm 0cm, clip=true, width=\columnwidth]{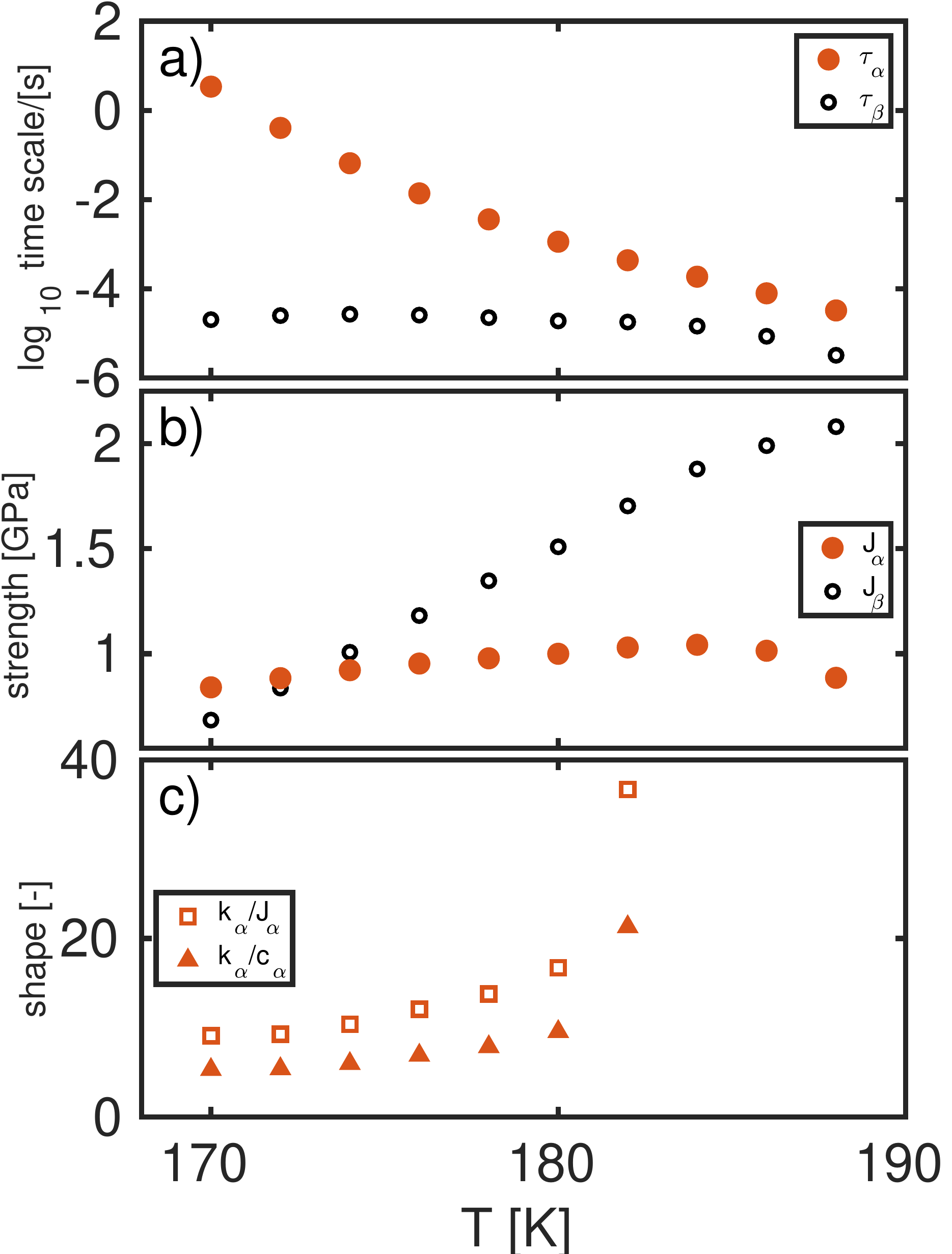}
	\caption{Fitting parameters for shear-mechanical data of Squalane.}
	\label{FigSQLNshearParams}
\end{figure}

\textbf{Shear-mechanical data} were fitted to the model applicable to dynamic shear-mechanical data as described by eq.~\ref{ABmodelShear}. It holds seven parameters to describe the response assuming a primary and secondary relaxation process (see Fig.~\ref{FigSQLNshearParams}). For the fitting, only one parameter was fixed to a constant (exponent of the $\beta$-relaxation, $b =$ 0.36), while the other six parameters were free.\\

\clearpage

\begin{figure}[h!]
\includegraphics[trim = 0cm 0cm 0cm 0cm, clip=true, width=\columnwidth]{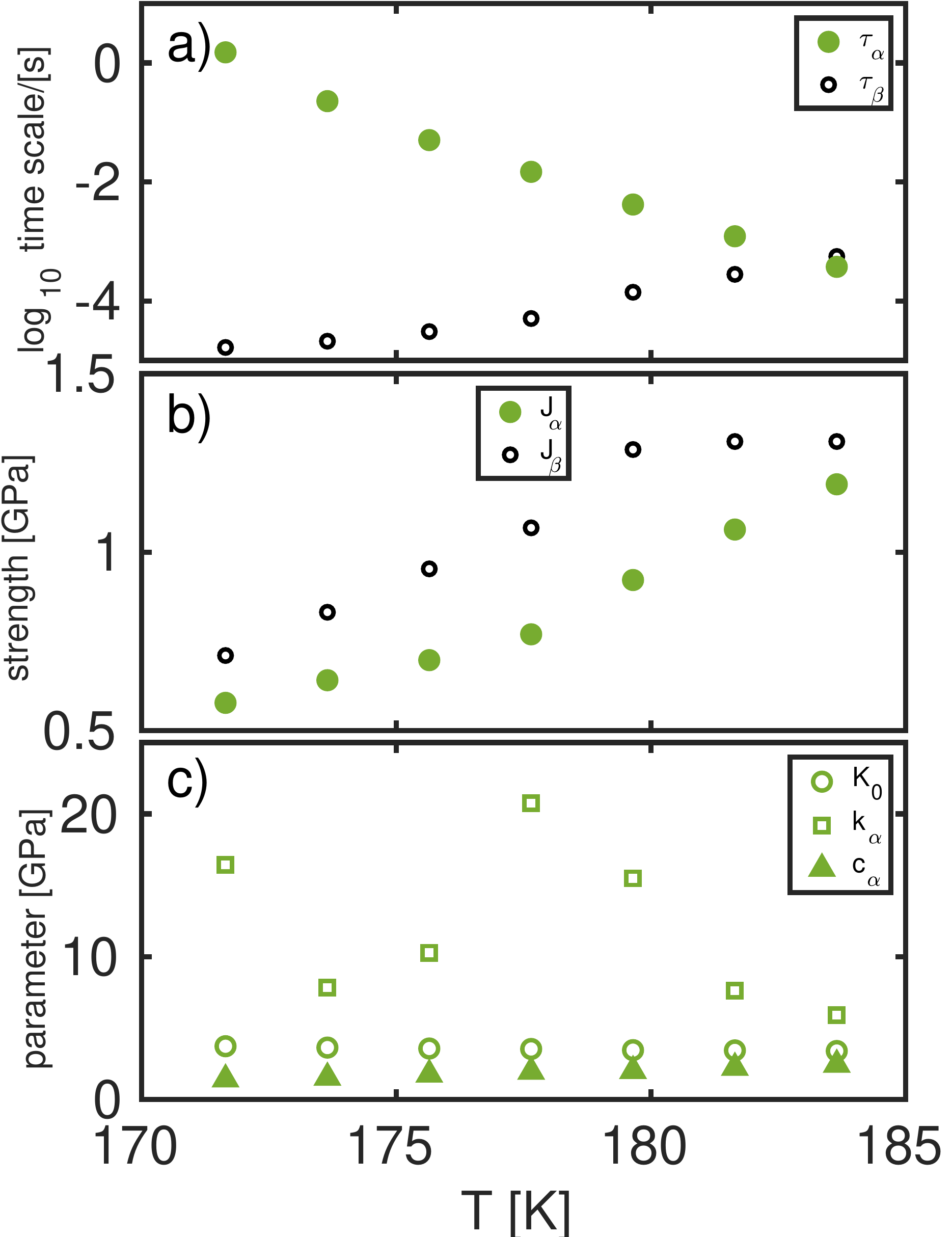}
	\caption{Fitting parameters for bulk-mechanical data of Squalane.}
	\label{FigSQLNbulkParams}
\end{figure}

\textbf{Bulk data} were fitted to the model applicable to dynamic bulk-mechanical data as described by eq.~\ref{ABmodelBulk}. It holds seven parameters to describe the response assuming a primary and secondary relaxation process. For the fitting, only one parameter was fixed to a constant (exponent of the $\beta$-relaxation, $b =$ 0.36), while the other parameters were free (see Fig.~\ref{FigSQLNbulkParams}).\\

\newpage

\begin{figure}[h!]
\includegraphics[trim = 0cm 0cm 0cm 0cm, clip=true, width=\columnwidth]{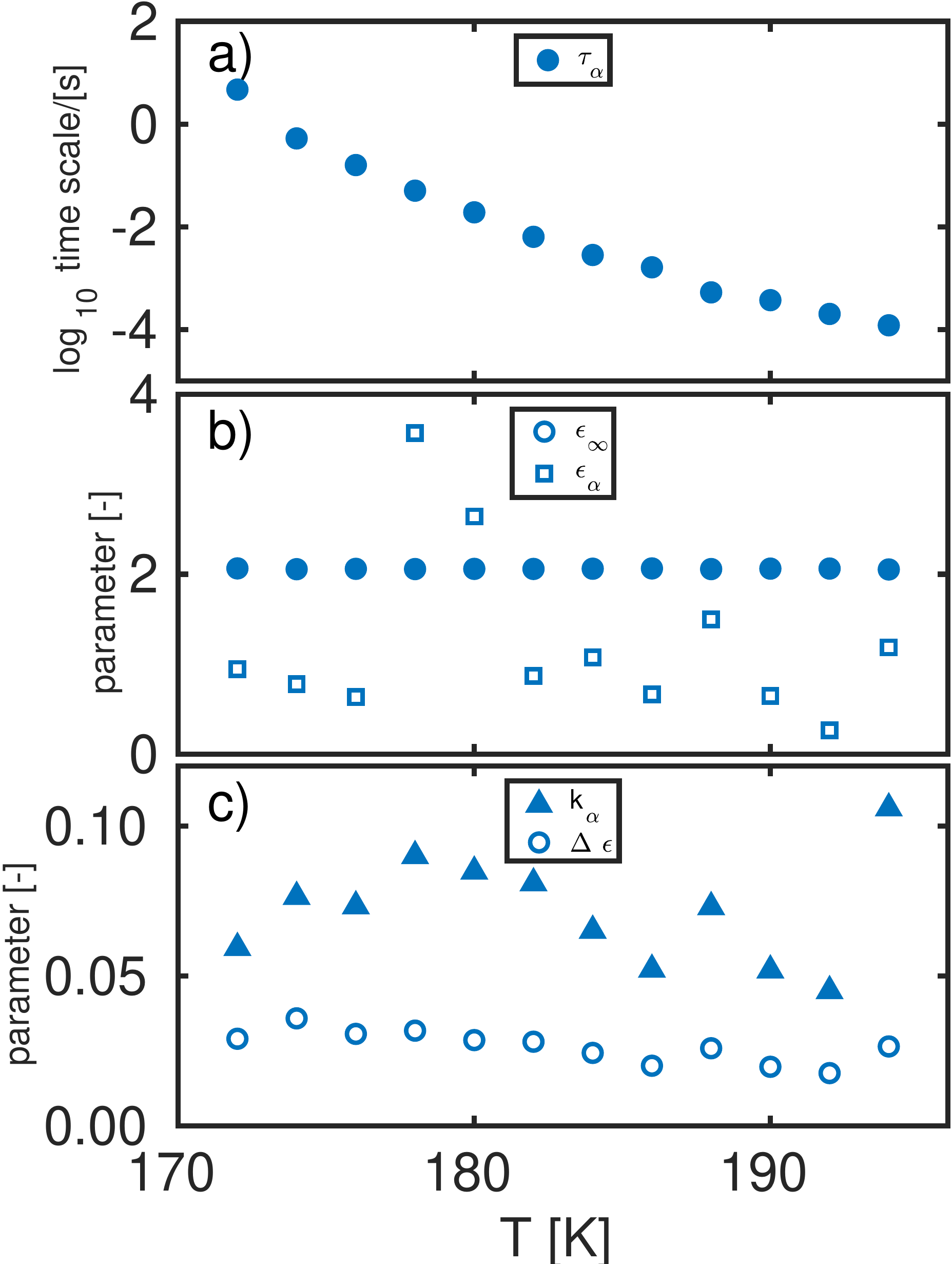}
	\caption{Fitting parameters for dielectric data of Squalane.}
	\label{FigSQLNdielParams}
\end{figure}

\textbf{Dielectric data} were fitted to the expression of the relevant EEC-model (eq.~\ref{ABmodelDiel3}). It holds seven parameters to describe the response assuming a primary and secondary relaxation process. For the fitting, only one parameter was fixed to a constant (exponent of the $\beta$-relaxation, $b =$ 0.36), while the other parameters were free (see Fig.~\ref{FigSQLNdielParams}).\\

\end{document}